\documentclass[fleqn,usenatbib,useAMS]{mnras}

\usepackage[T1]{fontenc}
\usepackage{ae,aecompl}
\usepackage[dvipsnames]{xcolor}

\usepackage{graphicx}	
\usepackage{amsmath}	
\usepackage{amssymb}	
\usepackage{natbib}

\title[Adapting a solid accretion scenario for migrating planets in FARGO3D]{Adapting a solid accretion scenario for migrating planets in FARGO3D}

\author[L. A. DePaula et al.]{
L. A. DePaula,$^{1}$\thanks{E-mail: luiz.paula@usp.br}
T. A. Michtchenko,$^{1}$
P. A. Sousa-Silva,$^{2}$
\\
$^{1}$Instituto Astron\^omico, Geof\'isico e Ci\^encia Atmosf\'ericas, Universidade de S\~ao Paulo, Rua do Mat\~ao 1226, 05508-900 S\~ao Paulo, Brazil\\
$^{2}$S\~ao Paulo State University - UNESP, Av. {Prof\textordfeminine}. Isette Corr\^ea Font\~ao, 505, 13876-750 - S\~ao Jo\~ao da Boa Vista, SP - Brazil
}

\date{Accepted 27/09/2019. Received 15/09/2019; in original form 21/06/2019}

\pubyear{2019}

\begin{document}
\label{firstpage}
\pagerange{\pageref{firstpage}--\pageref{lastpage}}
\maketitle

\begin{abstract}
In this work, we adapt a module for planetary formation within the hydrodynamic code FARGO3D. Planetary formation is modeled by a solid core accretion scenario, with the core growing in oligarchic regime. The initial superficial density of planetesimals is proportional to the initial superficial density of gas in the disc. We include a numerical approach to describe the evolution of the eccentricity and the inclination of planetesimals during the formation. This approach impacts directly on the accretion rate of solids. When the core reaches a critical mass, gas accretion begins, following the original FARGO scheme adapted to the FARGO3D code. To exemplify how the module for planetary formation can be used, we investigate the migration of a planet in a two-dimensional, locally isothermal gas disc with a prescribed accretion rate, analyzing the timescale involved in the planetary migration process along with the timescale for planetary formation. The analysis reveals that the mass of the nucleus must be close to its critical value when crossing the ice line to avoid the planet's fall into the stellar envelope. This will allow enough time for the planet to initiate runaway gas accretion, leading to a rapid mass increase and entering type II planetary migration. 
\end{abstract}




\section{Introduction}

According to the core accretion scenario, the formation of a giant planet begins when a solid nucleus forms by capturing planetesimals. After reaching a critical mass, the core is able to capture residual gas from the protoplanetary disc that will constitute its envelope, thus forming a giant planet \citep{Mizuno1980, Pollack1996}.

It is known that the gravitational interaction of a gas disc with a growing planet influences significantly its orbital parameters \citep{Papaloizou2007}. This happens when a sufficiently large planet (larger than Mars) produces spiral density waves in an initially axisymmetric  gas disc, breaking the symmetry and generating a resulting gravitational torque \citep{Goldreich1980, Ward1986}.
In general, in isothermal discs, a decay of the semi-major axis and a damping in the eccentricity occur \citep{Angelo2010}.
Planetary migration may occur by means of distinct processes. One is type I migration, during which energy and angular momentum of low mass planets are transferred to a gas disc, causing a rapid migration \citep{Ward1986,Ward1997,Papaloizou2007,Paardekooper2010}. Another is type II migration, when a massive planet opens a gap in a gas disc that decreases migration rate \citep{Bryden1999,Crida2009,Kley2012}. Type I migration is, in general, two orders of magnitude faster than type II migration \citep{Bitsch2010,Baruteau2014}. Additionally, there is type III migration regime in which the surface density distribution of gas in the co-orbital region is asymmetric, leading to a large torque and which is characterized by extremely short migration timescales \citep{Peplinski2008,Lin2010}. In general, type III migration applies to discs that are relatively massive and to planets that can only open partial gaps in the gas disc. 

An important issue when studying planetary formation/migration is to relate planetary formation and migration times to the lifetime of the disc \citep{Lissauer1993,Mizuno1980,Mordasini2010}. Numerical simulations indicate that time required to form a giant planet is longer than the lifetime of the disc. Besides, type I migration can be very rapid causing planets to fall in the stellar envelope before their complete formation \citep{Ida2008}. To avoid a planet's fall into the stellar envelope, one possible solution would be that the growing planet gains enough mass to enter type II migration regime. In this scenario, the ice line position in the disc plays an important role. Indeed, when a planet crosses this threshold during its inward migration, the superficial density of solids decreases drastically, making it difficult for the planet to gain enough mass to initiate gas accretion and enter type II migration \citep{Fortier2013}.

Many works concerning planetary formation use analytical developments to describe planetary migration \citep{Guilera2010,Fortier2013}. In general, for type I migration, analytical descriptions are limited to scenarios that are based on physically simple gas disc models, which allow hydrodynamic equations to be linearized \citep{Meyer1987, Tanaka2002}. The situation is more complicated when dealing with type II migration, since the planet's large mass creates a gap in the disc around the planet's orbit producing a nonlinearity and, consequently, making it very difficult to obtain an analytical prescription \citep{Bryden1999}. Moreover, analytical approaches are often inadequate when dealing with planetary migration due to several effects associated with the thermodynamics of the gas disc \citep{Paardekooper2010,Benitez2016}. Thus, a numerical approach using hydrodynamic simulators, such as FARGO3D \citep{Masset2000A, Benitez2016}, ZEUS \citep{Stone1992a}, PLUTO \citep{Mignone2012}, among others, becomes essential to study more realistic planetary migration scenarios, considering a wide range of physical conditions for the gas disc.

It is worth noting that few hydrodynamic simulators dealing with planet-gas disc interaction have models for planetary formation. Even when they have such models, in general, they only consider accretion of gas by giant planets \citep{Kley1999}, while not taking into account accretion of solids and gas for low-mass planets. This may be because the formation of planetary cores is usually modeled through N-body simulations, which, together with a hydrodynamic code, results in high computational costs.

In this work, we propose an approach that combines a statistical model for the solid core formation with FARGO3D hydrodynamic code, allowing us to significantly reduce computational costs, while reproducing results of N-body simulations. Specifically, the model for accretion of planetesimals is based on the works of \cite{Guilera2010} and \cite{Fortier2013}, which use a statistical model to determine the accretion rate of planetesimals \citep{Inaba2001}.

By implementing a planetary formation model in FARGO3D hydrodynamic simulator we are able to study, simultaneously, planetary formation and migration. In particular, in order to exemplify how the module for planetary formation can be used, we analyze the timescale involved in the migration process in conjunction with the timescale for planetary formation. This analysis reveals that, for the set of parameters chosen in this work, it is possible to obtain a planet's growth timescale shorter than the timescale for migration, even when the planet crosses the ice line during its inward migration. 
We show that, for planetesimals of radius $\sim$0.1\,km, it is possible to obtain planets that grow up to approximately 5 times the mass of Jupiter in regions between 0.5\,au and 1\,au in times shorter than the estimated lifetime of the gas disc ($\sim$10\,Myr).
This result was obtained using a fixed ice line position, defined by the values of temperature and pressure of the gas disc model in FARGO3D. 

We remark that, the ice line position can evolve with time \citep{Martin2012,Martin2013}. Also, it is known that accretion of peebles (solid material with sizes between cm and mm) has an important impact on planetary formation \citep{Lambrechts2014, Guilera2016, Johansen2017}. However, cores with a few Earth masses have a planetary envelope that could destroy these pebbles before they reach the nucleus \citep{Venturini2015}.  A model for peeble's accretion, for thermal effects and for the ice line evolution will be explored in future work.


In summary, this paper introduces a module for planetary growth model incorporated into FARGO3D hydrodynamic code. Our module comprises a statistical model for fixed-size planetesimals accretion and a model for gas accretion, allowing to investigate planetary migration along with planetary formation. 
To exemplify how the module for planetary formation can be used, we study the migration and formation of a planet in a two-dimensional isothermal disc with the ice line at a fixed position.

This paper is organized as follows. In Section 2 we describe the setup for the gas disc in FARGO3D hydrodynamic code. In Sections 3 and 4, we describe the models for solids and gas accretion, respectively, incorporated into FARGO3D. In Section 5, we discuss the results obtained, starting with the formation of a stationary solid core and continuing with the formation of a giant planet, also stationary. Following, we analyze the case in which the planet is allowed to migrate and discuss the relation between the position of the ice line and the critical mass of the solid core. In Section 6, we present the conclusions.

\section{Setup for the gas disc in FARGO3D}\label{setup}

In this work we use the hydrodynamic code FARGO3D \citep{Benitez2015}. We opt for this code version because it makes it possible to implement a model for planetary formation using GPU computing, thus reducing computational cost.

We use the same setup from \cite{DePaula2018}: we assume a thin two-dimensional gas disc (with vertical scale height, $H$, much smaller than its radius), locally isothermal and with viscosity given by the parameter $\alpha$ from \cite{Shakura1973}. The star, with $M_{\star} = 1\mathrm{M}_{\odot}$, is positioned at the origin of a cylindrical coordinate system, and the disc extends from 0.52\,au to 10.4\,au (0.1 to 2.0 in code units) \citep{DePaula2018}.

We use an equally spaced hydrodynamic grid with resolution of 582 $\times$ 1346 cells. To test if the chosen resolution is adequate, we start with a low resolution and increase it until the results obtained in Section~6 for the final mass and final position of a migrating planet differ in less than 5\%. Moreover, it is important that there are enough cells inside the Hill region of the planet during the runaway gas accretion regime to ensure consistent results in this phase. A test regarding the resolution during the gas accretion regime is included in Appendix \ref{app}.

It is worth mentioning that the influence of the resolution during planetary migration can be studied in further detail considering the number of cells in the half-horseshoe width region \citep{Paardekooper2010,Paardekooper2011}. This study could be especially important for migrating low mass planets. However, such investigation is beyond the scope of this paper, which is meant, mainly, to present the planetary formation module. Future work will deal with a moving mesh, so it will be possible to analyze more carefully the transition between migratory regimes during the planetary formation process.

As in \cite{Durman2015}, the radial component of the gas initial velocity is

\begin{equation}\label{eqvr2_1}
    v_{\mathrm{r}} = - \frac{3}{2} \alpha h^{2} r \Omega_{\mathrm{K}},
\end{equation}

\noindent{while its angular component is }

\begin{equation}\label{eqvr2_2}
    v_{\theta} = \sqrt{1 - \frac{3h^{2}}{2}} r \Omega_{\mathrm{K}},
\end{equation}

\noindent{where $h$ corresponds to the disc aspect ratio ($h = H/r = 0.05$)
and $\Omega_{\mathrm{K}} = \sqrt{GM_{\star}/r^{3}}$ is the Keplerian orbital frequency
of the gas at position $r$. The disc has an initial surface density of gas given by }

\begin{equation}\label{eqSigma}
    \Sigma_{\mathrm{gas}} = \frac{\dot{m}}{3\pi \alpha h^{2} \sqrt(G M_{\star})} r^{-1/2} = \Sigma_{0} r^{-1/2},
\end{equation}

\noindent{with $\Sigma_{0}$ denoting the surface density of gas at $r$ = 1 code units (corresponding to 5.2\,au) and $\dot{m}$ the gas disc accretion rate. Equations \ref{eqvr2_1}, \ref{eqvr2_2} and \ref{eqSigma} describe a stationary accretion disc with constant accretion rate \citep{Durman2015}. Following \cite{DePaula2018}, we use $\alpha$ = 0.003 and $\dot{m} = 10^{-7}$ $\mathrm{M}_{\odot}/\mathrm{year}$ to calculate $\Sigma_0$ for the disc's standard set up. Using equation \ref{eqSigma}, we obtain $\Sigma_{0}$ = 8743.4 $\mathrm{kg \cdot m^{-2}}$.}

We choose the same boundary conditions used in \cite{Durman2015}. As in \cite{DePaula2018}, we disable the damping of the surface density in FARGO3D to avoid gas accumulation at the inner boundary and allow gas to flow freely.
In this setup, the planet's initial mass is small. So, disc stabilization occurs very quickly. Particularly, we use a relaxation time of approximately 100 orbital periods for disc stabilization.

\section{Solid accretion rate: oligarchic growth regime}

\subsection{Initial surface density of solids and feeding zone}\label{feed}

Following \cite{Alibert2005,Fortier2013}, the initial superficial density of solids, $\Sigma_{\mathrm{m}}$, is assumed to be initially proportional to the initial surface density of gas, $\Sigma_{\mathrm{g}}$, that is,

\begin{equation}\label{densupsol01}
     \Sigma_{\mathrm{m}} = f_{\mathrm{D/G}} f_{\mathrm{R/I}} \Sigma_{\mathrm{g}} \ ,
\end{equation}

\noindent{where $f_{\mathrm{D/G}}$ is the gas and dust ratio in the disc, which, in general, scales with the star metallicity varying from 0.003 to 0.125 \citep{Fortier2013}. As per default we adopt $f_{\mathrm{D/G}}$ = 0.03, chosen to reduce computational time and to facilitate comparison with the literature. The factor $f_{\mathrm{R/I}}$ takes into account the degree of condensation of material due to the distance to the star. Following \cite{Fortier2013}, we adopt $f_{\mathrm{R/I}} = 1$ beyond the ice line, and $f_{\mathrm{R/I}} = 1/4$ inside the ice line.}
The position of the ice line depends on the height of the disc and the stellar radiation \citep{Min2011}. In our model, we choose to use a fixed position of 3.0 au, which was obtained by analyzing the standard disc temperature and pressure (see Section \ref{setup}). Also, we consider that the ice line does not evolve over time and there is no condensation nor sublimation of ice. These aspects are expected to be considered in future work.

As in \cite{Fortier2013}, we consider the disc of planetesimals to be composed by rocky and icy planetesimals with fixed size, assumed to be spherical with constant density. In particular, we assume that rocky planetesimals have a mean density of 3.2 $\mathrm{g/cm^{3}}$ and are located between the disc's innermost boundary and the ice line. 
On the other hand, we assume that icy planetesimals have a mean density of 1 $\mathrm{g/cm^{3}}$ and are located between the ice line and the disc's outermost boundary \citep{Fortier2013}. 
A model that considers planetesimals with different sizes \citep{Guilera2010} will be explored in future work.

Following \cite{Alibert2005}, planetesimals which can be accreted are those located in the protoplanet's feeding zone that consists of a ring around its orbit, with a width of $(b/2) R_{\mathrm{H}}$ on each side of the orbit, where $b$ is the width of the feeding zone, and $R_{\mathrm{H}} \approx a_{\mathrm{p}} \left( M_{\mathrm{p}}/{3M_{\star}} \right)^{1/3}$ is the Hill radius of the planet of mass $M_{\mathrm{p}}$. In this work, we adopt $b = 10$.

The surface density of planetesimals in the feeding zone changes as a result of the accretion onto the planet. Indeed, for each cell within the feeding zone, in each step $dt$ of the hydrodynamic code, the surface density of solids is reduced by

\begin{equation}\label{densupsol02}
    \Sigma_{\mathrm{m}} =  {\Sigma_{\mathrm{m}}}_{(0)} - \frac{\dot{M}_{\mathrm{core}} dt}{2 \pi a_{\mathrm{p}} b R_{\mathrm{H}}} \ ,
\end{equation}

\noindent{where ${\Sigma_{\mathrm{m}}}_{(0)}$ is the superficial density of solids in the cell,
$\dot{M}_{\mathrm{core}}$ is the accretion rate of solids and $a_{\mathrm{p}}$ is the distance from the protoplanet to the star.}

\subsection{The accretion rate of solids}

In our model for accretion of solids, a nucleus with initial mass $M_{\mathrm{core}}$ grows by accreting planetesimals of radius $r_{m}$ and mass $m$ in each step $dt$ of the hydrodynamic code. Adopting the particle-in-a-box approximation \citep{Chambers2006}, the solids accretion rate can be written as

\begin{equation}
   \frac{dM_{\mathrm{core}}}{dt} = \left( \frac{2 \pi \Sigma_{\mathrm{m}}}{P_{\mathrm{orbital}}}\right) P_{\mathrm{coll}},
\end{equation}

\noindent{where $\Sigma_{\mathrm{m}}$ is the mean azimuthal superficial solids density in the planet's feeding zone  (see Section \ref{feed}) and $P_{\mathrm{orbital}}$ is its orbital period. The collision rate $P_{\mathrm{coll}}$ corresponds the probability with which a planetesimal is accreted by the protoplanet. It depends on the relative velocity between the planetesimals and the protoplanet, which, in turn, depends on the planetesimals' eccentricities and inclinations \citep{Fortier2013}. In this work, $e$ and $i$ represent, respectively, the root mean square of the eccentricity and of the inclination of planetesimals inside the feeding zone. It is known that planetesimals are found in different velocity regimes, known as high, medium and low. Each regime leads to different collision rates. Applying \cite{Inaba2001}, the mean collision rate can be approximated by }

\begin{equation}
   P_{\mathrm{coll}} = \mathrm{min}(P_{\mathrm{med}}, (P_{\mathrm{high}}^{-2} + P_{\mathrm{low}}^{-2})^{-1/2}) \ ,
\end{equation}

\noindent{with}

\begin{equation}\label{phigh}
   P_{\mathrm{high}} = \frac{(R_{\mathrm{cap}} + r_{m})^2}{2 \pi R_{\mathrm{H}}} \left( F(\beta) + \frac{6 R_{\mathrm{H}} G(\beta)}{(R_{\mathrm{cap}} + r_{m})  {\tilde{e}}^2 } \right),
\end{equation}

\begin{equation}\label{pmed}
   P_{\mathrm{med}} = \frac{ (R_{\mathrm{cap}} + r_{m})^2 }{4 \pi {R_{\mathrm{H}}}^{2} {\tilde{i}}} \left( 17.3 + \frac{232 R_{\mathrm{H}}}{R_{\mathrm{cap}} + r_{m}}  \right),
\end{equation}

\begin{equation}\label{plow}
   P_{\mathrm{low}} = 11.3 \left( \frac{R_{\mathrm{cap}} + r_{m}}{R_{\mathrm{H}}} \right)^{1/2},
\end{equation}

\noindent{where $r_{m}$ is the radius of planetesimals, $R_{\mathrm{cap}}$ is the capture radius of the protoplanet (see Section \ref{accretiongas}), $\tilde{e} = a_{\mathrm{p}}e/R_{\mathrm{H}}$ is the reduced root mean square of the eccentricities, and $\tilde{i} = a_{\mathrm{p}}i/R_{\mathrm{H}}$ is the reduced root mean square of the inclinations. Moreover, $\beta = \tilde{i}/\tilde{e}$, with $F(\beta)$ and
$G(\beta)$ well-approximated by }

\begin{equation}
   F(\beta)  = \frac{1 + 0.95925\beta + 0.77251\beta^{2}}{\beta(0.13142 + 0.12295\beta)} ,
\end{equation}

\begin{equation}
   G(\beta)  = \frac{1 + 0.3996\beta}{\beta(0.0369 + 0.0048333\beta + 0.006874\beta^{2})}  ,
\end{equation}

\noindent{for $0 < \beta \leq 1$, which is the range of interest for this work \citep{Chambers2006}.}

The solid core grows until all material is depleted from within the feeding zone or until it reaches a mass $M_{\mathrm{sca}}$ for which planetesimals are ejected. Following \cite{Ida2004a}, we use

\begin{equation}\label{eqmscat}
   M_{\mathrm{sca}} = \frac{2 R_{\mathrm{core}} f_{\mathrm{cap}} M_{\star}}{a_{\mathrm{p}}}  ,
\end{equation}

\noindent{where $R_{\mathrm{core}}$ is the geometric radius of the core and $f_{\mathrm{cap}} = 0.1$.}

\subsection{Evolution of the root mean square eccentricity and inclination}\label{evroot}

Equations \ref{phigh}, \ref{pmed} and \ref{plow} depend on the root mean square of the eccentricity ($e$) and of the inclination $(i)$ of planetesimals inside the feeding zone. Following \cite{Fortier2013}, we consider two different cases for the evolution of $e$ and $i$.

In the first case, called equilibrium scenario, $e$ and $i$ are estimated assuming the gas drag instantaneously balances the stirring due to the protoplanet \citep{Thommes2003}. Thus,

\begin{equation}\label{eqecceq}
   e = 2i = 1.7 \frac{ m^{1/15} {M_{\mathrm{core}}}^{1/3} \rho_{m}^{2/15} }{ b^{1/5} \mathrm{C}_{\mathrm{D}}^{1/5} \rho_{\mathrm{gas}}^{1/5} M_{\star}^{1/3} a_{\mathrm{p}}^{1/5} },
\end{equation}

\noindent{where $b$ is the width of the feeding zone, $m$ is the mass of the planetesimal, $M_{\mathrm{core}}$ is the mass of the solid core, $\rho_{m} = 3.2$ $\mathrm{g/cm^{3}}$ is the bulk density of the planetesimal 
before the ice line (after the ice line, $\rho_{m} =1.0$ $\mathrm{g/cm^{3}}$), $\rho_{\mathrm{gas}}$ is the volumetric density of gas and $\mathrm{C}_{\mathrm{D}}$ is the drag coefficient, which is of the order of one.} 

Since we perform a 2D simulation within the 3D FARGO3D code, the volumetric density of gas is not computed directly. We replace its value by an estimate which results from integrating the hydrostatic equilibrium equation obtained for an isothermal disc \citep{Armitage2010}, such that

\begin{equation}\label{analpad01}
   \rho_{\mathrm{gas}} = \frac{1}{\sqrt{2 \pi}} \frac{\Sigma_{\mathrm{gas}}}{hr} \ ,
\end{equation}

\noindent{where $\Sigma_{\mathrm{gas}}$ is the surface density of gas at $r$.}

In the second case, called the out of equilibrium scenario, the eccentricities and inclinations evolve according to 

\begin{equation}\label{eqtodas1}
   \frac{d {e_{ij}}^{2}}{dt} = \left( \frac{d {e_{ij}}^{2}}{dt} \right)_{\mathrm{gas}} + \left( \frac{d {e_{ij}}^{2}}{dt} \right)_{\mathrm{grav}},
\end{equation}

\begin{equation}\label{eqtodas2}
   \frac{d {i_{ij}}^{2}}{dt} =  \left( \frac{d {i_{ij}}^{2}}{dt} \right)_{\mathrm{gas}} + \left( \frac{d {i_{ij}}^{2}}{dt} \right)_{\mathrm{grav}},
\end{equation}

\noindent{where the first terms represent the gas drag and the second terms are the viscous stirring produced by an embryo of mass $M_{\mathrm{core}}$, with subscripts $i$ and $j$ refering to the cell position. Therefore, in this case,  specific values for the eccentricity and the inclination are obtained for each cell. Then, the root mean squares of eccentricities and of inclinations ($e$ and $i$) in the feeding zone are calculated. In this work, we disregard interactions between planetesimals because they do not affect significantly how $e$ and $i$ evolve \citep{Fortier2013}.}

The drag force experienced by a spherical body depends on the relative velocity with respect to the gas and the mean free path of a molecule of gas, $\lambda = (n_{{\mathrm{H}_{2}}} \sigma_{{\mathrm{H}_{2}}})^{-1}$. There are three different regimes corresponding to how drag force acts on a planetesimal. First and second regimes, known as quadratic and Stokes regime, respectively, occur for planetesimals with radius larger than the mean free path, $r_{m} \gtrsim \lambda$. 
The criterion proposed by \cite{Rafikov2004} is used to identify these two regimes. This criterion characterizes these regimes using the molecular Reynold number $\mathrm{Re_{mol}} = v_{\mathrm{rel}} r_{m} / \nu_{\mathrm{gas}} $, where $\nu_{\mathrm{gas}} = \lambda c_{\mathrm{s}}/3$, $v_{\mathrm{rel}} = v_{\mathrm{k}} \sqrt{5/9 e_{ij}^{2} + 1/2 i_{ij}^{2}}$ ($v_{\mathrm{k}}$ is the Keplerian velocity in the cell) and $c_{\mathrm{s}}$ is the velocity of the sound in the cell.

For $\mathrm{Re}_{\mathrm{mol}} > 20$ and $r_{m} > \lambda$, the quadratic regime occurs. Thus

\begin{equation}
\frac{d e_{ij}^{2}}{dt} \bigg|_{\mathrm{drag}} = -\frac{2e_{ij}^{2}}{\tau_{\mathrm{drag}}} \left( \frac{9}{4} \eta^{2} + \frac{9}{4\pi} \xi^{2}e_{ij}^{2} + \frac{1}{\pi} i_{ij}^{2} \right) ,
\end{equation}

\begin{equation}
\frac{d i_{ij}^{2}}{dt} \bigg|_{\mathrm{drag}} = -\frac{i_{ij}^{2}}{\tau_{\mathrm{drag}}} \left( \eta^{2} + \frac{1}{\pi} \xi^{2}e_{ij}^{2} + \frac{4}{\pi} i_{ij}^{2} \right),
\end{equation}

\noindent{where $\xi = 1.211$ \citep{Inaba2001}, $\eta$ is given by \citep{Takeuchi2002}, }

\begin{equation}
\eta =  \left( \frac{h}{r} \right)^{2} (p + q + \gamma \frac{z}{h}),
\end{equation}

\noindent{where $p = \varphi$ (two-dimensional disc) and $q = 2.0 \gamma - 3.0$ , with $ \gamma = 0$ defining the disc curvature and $\varphi = 1/2$ giving the initial surface density curve (see Eq. \ref{eqSigma}).}

Besides that, $\tau_{\mathrm{drag}}$ is given by \citep{Chambers2006}:

\begin{equation}
\tau_{\mathrm{drag}} = \frac{8 \rho_{m}r_{m}}{3 \mathrm{C}_{\mathrm{D}} \rho_{\mathrm{gas}} v_{\mathrm{K}}}.
\end{equation}

On the other hand, for $\mathrm{Re}_{\mathrm{mol}} < 20$ and $r_{m} > \lambda$, Stokes regime occurs. Then

\begin{equation}
\frac{d e_{ij}^{2}}{dt} \bigg|_{\mathrm{drag}} = -\frac{3}{2} \frac{\lambda c_{\mathrm{s}} \rho_{\mathrm{gas}} e_{ij}^{2}}{\rho_{\mathrm{m}} r_{m}^{2}},
\end{equation}

\begin{equation}
\frac{d i_{ij}^{2}}{dt} \bigg|_{\mathrm{drag}} = -\frac{3}{4} \frac{\lambda c_{\mathrm{s}} \rho_{\mathrm{gas}} i_{ij}^{2}}{\rho_{\mathrm{m}} r_{m}^{2}}.
\end{equation}

Finally, when $r_{m} < \lambda$, a third regime, called Epstein regime, takes place. In this case

\begin{equation}
\frac{d e_{ij}^{2}}{dt} \bigg|_{\mathrm{drag}} = -e_{ij}^{2} \frac{c_{\mathrm{s}} \rho_{\mathrm{gas}}}{\rho_{\mathrm{m}} r_{m}},
\end{equation}

\begin{equation}
\frac{d i_{ij}^{2}}{dt} \bigg|_{\mathrm{drag}} = -\frac{i_{ij}^{2}}{2} \frac{c_{\mathrm{s}} \rho_{\mathrm{gas}}}{\rho_{\mathrm{m}} r_{m}}.
\end{equation}

The eccentricities and inclinations of planetesimals are also influenced by the protoplanet's gravitational potencial. According to \cite{Ohtsuki2002}, this effect is described by

\begin{equation}
\frac{d e_{ij}^{2}}{dt} \bigg|_{\mathrm{VS,M}} = \left( \frac{M_{\mathrm{core}}}{3bM_{\star} P_{\mathrm{orbital}}} \right) P_{\mathrm{VS}} ,
\end{equation}

\begin{equation}
\frac{d i_{ij}^{2}}{dt} \bigg|_{\mathrm{VS,M}} = \left( \frac{M_{\mathrm{core}}}{3bM_{\star} P_{\mathrm{orbital}}} \right) Q_{\mathrm{VS}} ,
\end{equation}

\noindent{where $P_{\mathrm{VS}}$ and $Q_{\mathrm{VS}}$ are functions of $\tilde{e}_{ij}$, $\tilde{i}_{ij}$ and $\beta_{ij}$,
and are given by }

\begin{equation}\label{eveccinc09}
   \begin{cases}
   \begin{split}
   \displaystyle{  P_{\mathrm{VS}} = \left[ \frac{73 \tilde{e}_{ij}^2}{10 \Lambda^2} \right] \ln{(1 + 10\Lambda^2 / \tilde{e}_{ij}^2)} +} & \\
   \displaystyle{ \left[ \frac{72 I_{\mathrm{PVS}} (\beta)}{\pi \tilde{e}_{ij} \tilde{i}_{ij} } \right] \ln{(1 + \Lambda^2)} } \ , & \\
   \displaystyle{ Q_{\mathrm{VS}} = \left[ \frac{4 \tilde{i}_{ij}^2 + 0.2 \tilde{i}_{ij} \tilde{e}_{ij}^3}{10 \Lambda^2 \tilde{e}_{ij}} \right] \ln{(1 + 10\Lambda^2 \tilde{e}_{ij})} +} & \\
   \displaystyle{ \left[ \frac{72 I_{\mathrm{QVS}} (\beta)}{\pi \tilde{e}_{ij} \tilde{i}_{ij} } \right] \ln{(1 + \Lambda^2)} }, &
   \end{split}
   \end{cases}
\end{equation}

\noindent{with $\Lambda = \tilde{i}_{ij} (\tilde{e}_{ij}^{2} + \tilde{i}_{ij}^{2})/12$. Functions $I_{\mathrm{PVS}}$ and $I_{\mathrm{QVS}}$ can be approximated to $0 <\beta_{ij} \leq 1$ by }

\begin{equation}\label{eveccinc10}
   \begin{cases} \displaystyle{ I_{\mathrm{PVS}} (\beta_{ij}) \simeq \frac{\beta_{ij} - 0.36251}{0.061547 + 0.16112\beta_{ij} + 0.054473 \beta_{ij}^2} } \ , \\
                 \displaystyle{ I_{\mathrm{QVS}} (\beta_{ij}) \simeq \frac{0.71946 - \beta_{ij}}{0.21239 + 0.49764\beta_{ij} + 0.14369 \beta_{ij}^2} } \ ,
   \end{cases}
\end{equation}

\noindent{where $\beta_{ij} = \tilde{i}_{ij}/\tilde{e}_{ij}$.}

As the distances between protoplanet and planetesimals increase, the excitation produced on planetesimals' eccentricities and inclinations decreases. Following \cite{Guilera2010}, we use

\begin{equation}\label{eveccinc11}
   \begin{cases} \displaystyle{ \left( \frac{d e_{ij}^{2}}{dt} \right)_{\mathrm{grav}}^{\mathrm{eff}} = f(\Delta) \left( \frac{d e_{ij}^{2}}{dt} \right)_{\mathrm{grav}}  } \ , \\
                 \displaystyle{ \left( \frac{d i_{ij}^{2}}{dt} \right)_{\mathrm{grav}}^{\mathrm{eff}} = f(\Delta) \left( \frac{d i_{ij}^{2}}{dt} \right)_{\mathrm{grav}}  } \ ,
   \end{cases}
\end{equation}

\noindent{where $f(\Delta)$ ensures that the perturbation is bounded within a neighborhood of the protoplanet and is given by }

\begin{equation}\label{eveccinc12}
   f(\Delta) = \left[ 1 + \left( \frac{\Delta}{n R_{\mathrm{H}}} \right) \right] \ .
\end{equation}

In this work, we adopt $n = 5$ and we define $\Delta$ as the radial distance between the center of the cell containing the protoplanet and the center of the cell used to calculate evolution \citep{Guilera2010}. 

To solve equations \ref{eqtodas1} and \ref{eqtodas2}, we need to specify initial conditions for $e_{0}$ and $i_{0}$. Following \cite{Fortier2013}, we consider two possibilities: the first set of initial conditions is given by the value corresponding to equilibrium between gas drag and mutual stirring on the planetesimals, such that

\begin{equation}\label{ecctipo2}
   e_{ij} = 2i_{ij} = 2.31 \frac{m^{4/15} \Sigma_{\mathrm{gas}}^{1/5} a_{\mathrm{p}}^{1/5} \rho_{m}^{2/15}}{\mathrm{C}_{\mathrm{D}}^{1/5} \rho_{\mathrm{gas}}^{1/5} M_{\star}^{2/5}} \,
\end{equation}

The second set of initial conditions assumes that the planetesimals are already excited by the embryo and their initial eccentricities and inclinations are given by equation \ref{eqecceq}.

Table \ref{table1} summarizes the initial values $e$ in each case, for planetesimals with $r_{m} = 100$, 10, 1, and 0.1~km.

\begin{table}
\caption{Initial values of $e$ obtained using Eq.~\ref{eqecceq} and Eq.~\ref{ecctipo2} for each size of planetesimals}             
\label{table1}      
\centering                          
\begin{tabular}{c c c}        
\hline\hline                 
$r_{m}$ & Eq.~\ref{eqecceq} & Eq.~\ref{ecctipo2}  \\    
\hline                        
   0.1 km & 0.0050 & 0.00010\\      
   1 km & 0.0079 & 0.00065\\
   10 km  & 0.013  & 0.0041\\
   100 km  & 0.020  & 0.025\\
\hline                                   
\end{tabular}
\end{table}

\begin{figure*}
\resizebox{\hsize}{!}
{\includegraphics[width=\hsize]{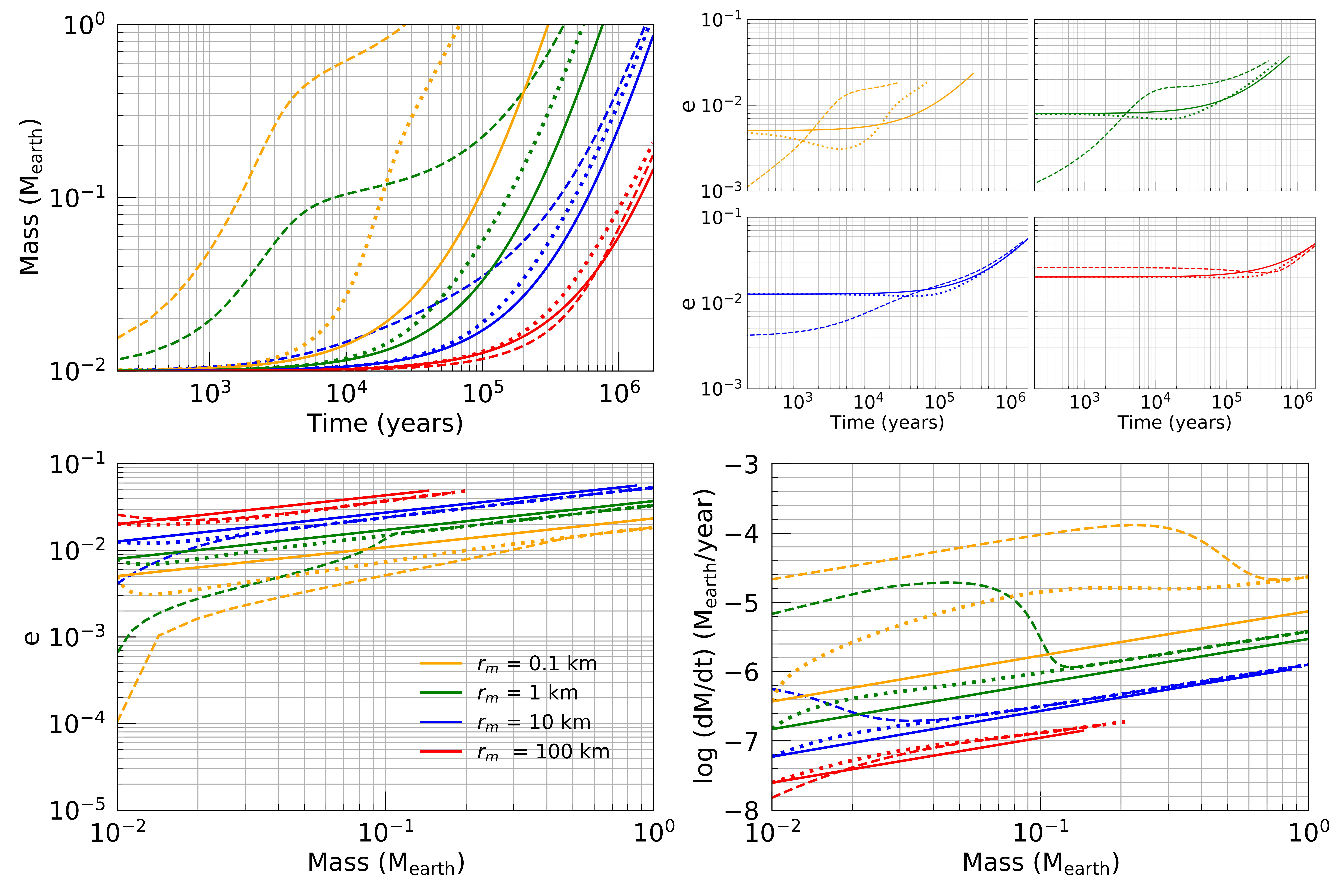}}
\caption{Formation of a solid embryo at 6\,au, without migration. Colors represent planetesimals with different radii (red, 100 km; blue, 10 km; green, 1 km; orange, 0,1 km). The embryo's initial mass is 0.01$\mathrm{M}_{\mathrm{earth}}$; it is allowed to grow either until it reaches 1$\mathrm{M}_{\mathrm{earth}}$ or until time reaches $1.8 \times 10^{6}$ years. The solid line corresponds to the equilibrium scenario (see Eq. \ref{eqecceq}), while the dashed and dotted lines correspond to the out of equilibrium scenario for two different initial conditions (Eq. \ref{ecctipo2} for the dashed line and Eq. \ref{eqecceq} for the dotted line). The top-left panel shows the planet's mass as a function of time. In the top-right panel, the time evolution of $e$ is shown. The bottom-left panel shows the same quantity as a function of the planet's growing mass. The bottom-right panel shows how the accretion rate of planetesimals evolves with the mass of the planet.}
\label{resform1MT}
\end{figure*}

\section{Gas accretion rate}\label{accretiongas}

When the core reaches a critical mass, $M_{\mathrm{crit}}^{\mathrm{core}}$, gas accretion begins. A set of numerical solutions, based on the physical structure of the envelope in quasi static equilibrium \citep{Ikoma2000}, gives an estimation for the core's critical mass

\begin{equation}\label{formgigantes1}
  M_{\mathrm{crit}}^{\mathrm{core}} \simeq 10 \left( \frac{\dot{M}_{\mathrm{core}}}{10^{-6} \ \mathrm{M}_{\mathrm{earth}} \mathrm{yr^{-1}}} \right)^{c} \left( \frac{\kappa}{ 1 \ \mathrm{ cm^{2} g^{-1}}} \right)^{c} \mathrm{M}_{\mathrm{earth}} \ ,
\end{equation}

\noindent{where $\dot{M}_{\mathrm{core}}$ is the rate of solids accretion and $\kappa$ is the grain opacity in the outermost envelope.}

Following \cite{Ida2004b}, we adopt $c = 0.25$ for the power law in $\dot{M}_{\mathrm{core}}$, and, considering an abundance of dust particles similar to the one in the interstellar medium, we use $\kappa \sim 1$ $\mathrm{cm}^{2} \mathrm{g}^{-1}$. This opacity value is debatable, since the fraction of elements during this phase of planetary formation is not very well known.

Once $M_{\mathrm{crit}}^{\mathrm{core}}$ is obtained, the code computes the gas accretion rate as the minimum between the potential value of the gas accretion rate ($\dot{M}_{\mathrm{U}}$) and the Kelvin-Helmholtz gas accretion rate ($\dot{M}_{\mathrm{KH}}$), for a particular time step $dt$ \citep{DePaula2018}. For a comprehensive description of the gas accretion model used in this work, including how to compute $\dot{M}_{\mathrm{U}}$ and $\dot{M}_{\mathrm{KH}}$, we refer the reader to \cite{DePaula2018} which deals exclusively with the gas accretion model used here. 

We also consider the capture radius for solids accretion, which has a significant influence on the accretion rate. In fact, the capture radius takes into account the planetary envelope, which reduces planetesimals' velocities and, consequently, increases the domain from which planetesimals can be accreted. In this work, we follow the model of \cite{Ormel2012}, in which a low opacity radiative envelope and an approximation for the planetary envelope structure are used. The capture radius, $R_{\mathrm{cap}}$, is given by

\begin{equation}
R_{\mathrm{cap}} \approx R_{\mathrm{B}}
   \begin{cases}
   \begin{split}
   \displaystyle{\left[ 1 + \frac{2 W_{\mathrm{neb}} (\sigma_{a} - 1) + \log{\sigma_{a}}}{\gamma} \right]^{-1}} \ , & \\
   \displaystyle{(1 \leqslant \sigma_{a} \leqslant \sigma_{1})} \ , & \\
   \displaystyle{\left[ \frac{1}{x_{1}} + \frac{4 (W_{\mathrm{neb}})^{1/3}}{\gamma} (\sigma_{a}^{1/3} - \sigma_{1}^{1/3}) \right]^{-1}} \ , & \\
   \displaystyle{(\sigma_{a} > \sigma_{1})}, &
   \end{split}
   \end{cases}
\end{equation}

\noindent{where $R_{\mathrm{B}} = G M_{\mathrm{p}}/\gamma c_{\mathrm{s}}^{2}$ (with $\gamma = 1.4$) is the Bondi radius of the planet with the mass $M_{\mathrm{p}}$ and $\sigma_{a}$ is the ratio between the density required for capturing planetesimals and the volumetric density of gas. According to \cite{Ormel2012}, we can write }

\begin{equation}
  \sigma_{a} = \frac{(6 + e^{2}) r_{m} \rho_{\mathrm{m}}}{9 R_{\mathrm{H}}} ,
\end{equation}

\noindent{where $r_{m}$ and $\rho_{\mathrm{m}}$ are the planetesimals' radius and volumetric density, respectively.}

The magnitude $\sigma_{1} = 1/(5 W_{\mathrm{neb}})$ is a dimensionless quantity defining a transition between gas pressure regimes \citep{Ormel2012} and $x_{1} = R_{1}/R_{\mathrm{B}} = 1 + 2 W_{\mathrm{neb}} (\sigma_{1} - 1) + \log{\sigma_{1}}$ is a normalization parameter of the radius of the envelope. The envelope's atmosphere structure is defined by parameter $W_{\mathrm{neb}}$, given by

\begin{equation}
  W_{\mathrm{neb}} = \frac{3 \kappa_{\mathrm{B}} L_{c}}{64 \pi \sigma_{\mathrm{sb}}} \frac{P_{\mathrm{gas}}}{G M_{\mathrm{core}} T_{\mathrm{gas}}^{4}} ,
\end{equation}

\noindent{where $\kappa_{\mathrm{B}} = 0.01$ $\mathrm{cm}^{2} \cdot \mathrm{g}^{-1}$ is the envelope's internal opacity, $\sigma_{\mathrm{sb}}$ is the Stefan-Boltzman constant, $P_{\mathrm{gas}}$ is the gas pressure, $T_{\mathrm{gas}}$ is the gas temperature and $L_{c}$ is the luminosity due to the accretion of planetesimals, which is given by}

\begin{equation}
  L_{c} = \left( \frac{G M_{\mathrm{core}}}{R_{\mathrm{core}}} \right) \dot{M}_{\mathrm{core}},
\end{equation}

\noindent{where $R_{\mathrm{core}}$ is the geometric radius of the core.}

We remark that, in this work, we consider a simplification about the opacity. During the envelope's formation, we use a low opacity ($\kappa_{\mathrm{B}} = 0.01$ $\mathrm{cm}^{2} \cdot \mathrm{g}^{-1}$) to compute the capture radius for accretion of solids. On the other hand, to compute the critical mass ($\kappa \sim 1$ $\mathrm{cm}^{2} \mathrm{g}^{-1}$), we consider the opacity in the outermost envelope. This simplification, while debatable, is needed to deal with how complex this topic is. For further details, we refer the reader to \cite{Movshovitz2008}.

\section{Results}

\subsection{Formation of a solid embryo}\label{solidformation}

In this section, we focus on the initial stages of solids accretion. We are interested in assessing how the size of accreted planetesimals affects the growth of an embryo. The results obtained are compared with those obtained by \cite{Fortier2013}.

We analyze the formation of a solid embryo with initial mass of 0.01$\mathrm{M}_{\mathrm{earth}}$ and position fixed at 6\,au, that is, a planet which is not allowed to migrate. The initial density of solids is set to 244\,$\mathrm{kg}/\mathrm{m}^{2}$ (see Sec. \ref{feed}).  Figure \ref{resform1MT} presents some results. The top-left panel shows the mass of the planet as a function of time. The top-right panel shows how $e$ evolves as a function of time, while the bottom-left panel shows the same quantity as a function of the planet's mass. Finally, the bottom-right panel shows the accretion rate of planetesimals as a function of the planet's mass. Simulation stops when either the mass reaches 1$\mathrm{M}_{\mathrm{earth}}$ or the formation time reaches $1.8 \times 10^{6}$ years. On all panels, the families of curves are parameterized by the  planetesimals' radii, associated with the colors shown on the left-bottom panel.

In order to easily compare these results with \cite{Fortier2013}, we use similar line patterns on all panels of Fig. \ref{resform1MT}. Solid lines represent the equilibrium scenario, in which $e$ and $i$ evolve according to Equation \ref{eqecceq}. For the out of equilibrium scenario, we solve Equations \ref{eqtodas1} and \ref{eqtodas2} using two different sets of initial conditions for the eccentricity and inclination of planetesimals, as done in \cite{Fortier2013}. Dashed lines illustrate the scenario corresponding to initial conditions given by Equations \ref{ecctipo2}, while dotted lines describe the scenario obtained for initial conditions given by Equation \ref{eqecceq}. The second set of initial conditions describes a hot disc, in which the eccentricities and inclinations have already been excited by the embryo \citep{Fortier2013}.

On the top-left panel of Fig. \ref{resform1MT}, we note that the time required to form a solid nucleus with about one Earth mass is much shorter for small planetesimals than for larger ones. Indeed, for planetesimals with 100 km of radius, a time of $1.8 \times 10^{6}$ years is not sufficient to form a nucleus of one Earth mass. On the other hand, for planetesimals with 0.1 km of radius, a nucleus with one Earth mass is formed in less than $10^{5}$ years. The same behavior was described by \cite{Fortier2013}; it occurs because smaller planetesimals are more affected by gas drag. In fact, the top-right panel in Fig. \ref{resform1MT} shows that, when planetesimals are large, $e$ remains high during the growth process, making accretion process much slower.

Comparing our results with those in \cite{Fortier2013}, we obtain, in general, shorter times to form a solid nucleus, especially with small planetesimals. This is essentially because each work uses different models for the gas disc and because we use a larger value for the gas and dust ratio ($f_{\mathrm{D/G}} = 0.03$) in order to reduce computational cost. However, it is worth noticing that, setting the same parameters provided in \cite{Fortier2013}, we reproduce their results, thus validating our model.

The top-left panel of Fig. \ref{resform1MT} shows that the equilibrium and the out of equilibrium scenarios require 
different times to form a solid nucleus of one Earth mass, especially when considering small planetesimals. For example, for planetesimals of 0.1 km radius, formation time differs by approximately $\simeq 1.8 \times 10^{5}$ years, while for planetesimals of 1 km radius this difference is of about $\simeq 2.2 \times 10^{5}$ years. This behaviour is also found by \cite{Fortier2013}.
We also note that, especially with small planetesimals, the evolution of $e$ (Fig. \ref{resform1MT}, top-right) has a strong impact on the rate of accretion of planetesimals (Fig. \ref{resform1MT}, lower-right).

\begin{figure}
\centering
\includegraphics[width=1.0 \columnwidth,angle=0]{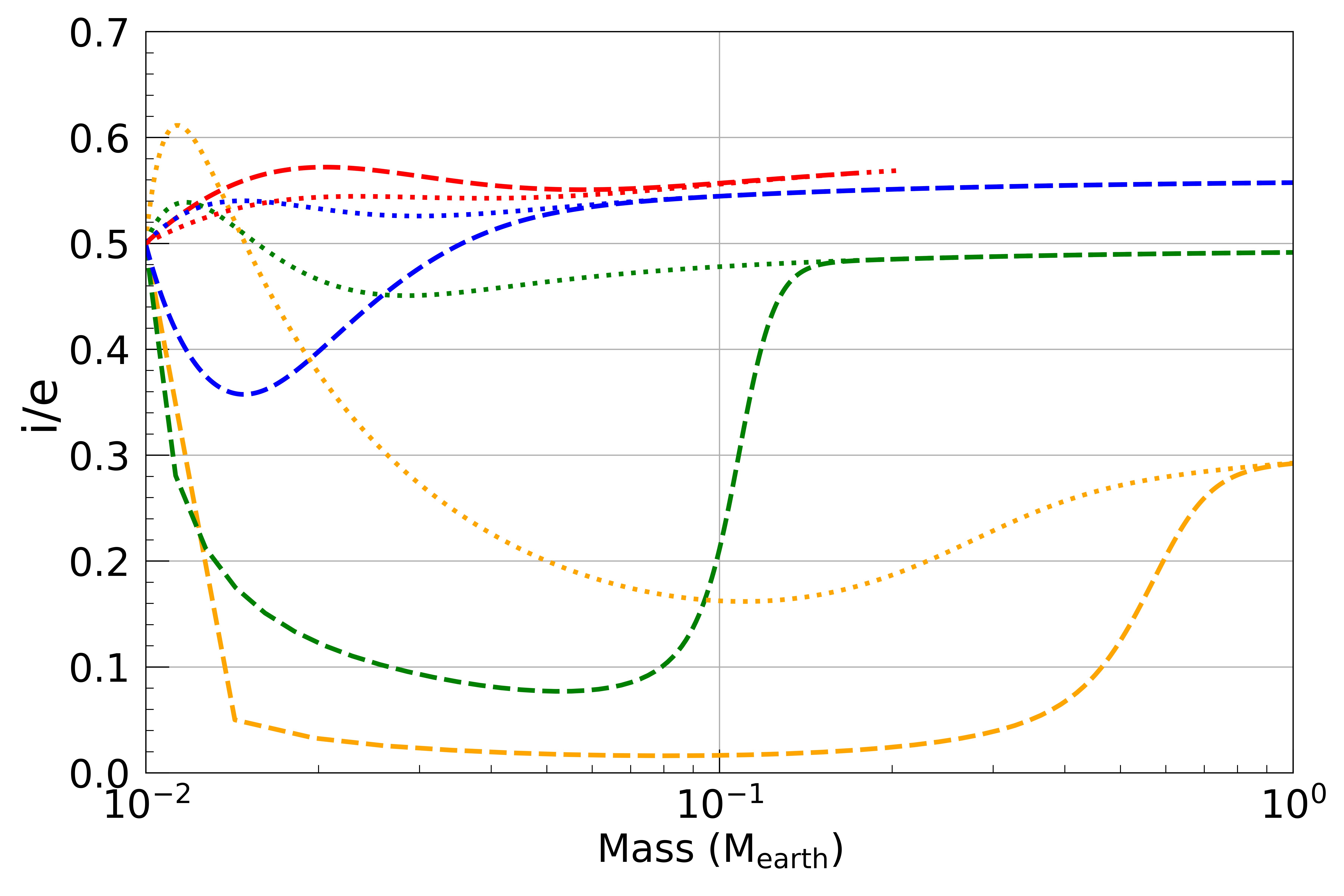}
\caption{Evolution of the ratio between the root mean square of the inclinations and eccentricities in the feeding zone in the out of equilibrium scenario. Colors represent planetesimals with different radii (red, 100 km; blue, 10 km; green, 1 km; orange, 0,1 km). Dashed and dotted lines correspond to the out of equilibrium scenario for two different initial conditions given by Eq. \ref{ecctipo2}, (dashed line) and Eq. \ref{eqecceq} (dotted line).}
\label{fargoformdivxmass}
\end{figure}

Figure \ref{fargoformdivxmass} shows the evolution of the ratio between $i$ and $e$ in the out of equilibrium scenarios during the solid core growth. We see that, for small planetesimals, this ratio is very different from $i/e = 0.5$ in the equilibrium scenario. Then, we conclude that the equilibrium scenario works well for large planetesimals, but the time to form a massive nucleus, in this case, is very high and the formation of giant planets is compromised.

In general, our results are in good agreement with those obtained with 1D disc models \citep{Guilera2010, Fortier2013}. However, it is worth noting that we developed a 2D disc model (that could be expanded to 3D), thus, our approach can be understood as a first step in the direction of a more complex modeling.

\subsection{Formation of a giant planet}\label{solidgasformation}

In this section, we analyze the formation of a giant planet, following the same procedure of Section \ref{solidformation}, with the same distribution of solids and the same initial conditions for the gas disc. However, due to the high computational costs, we choose a planet with larger initial mass, especifically, 0.1$\mathrm{M}_{\mathrm{earth}}$, fixed at 6\,au, which accretes solids until the critical mass is reached (see Eq. \ref{formgigantes1}), and gas accretion begins.

\begin{figure*}
\resizebox{\hsize}{!}
{\includegraphics[width=\hsize]{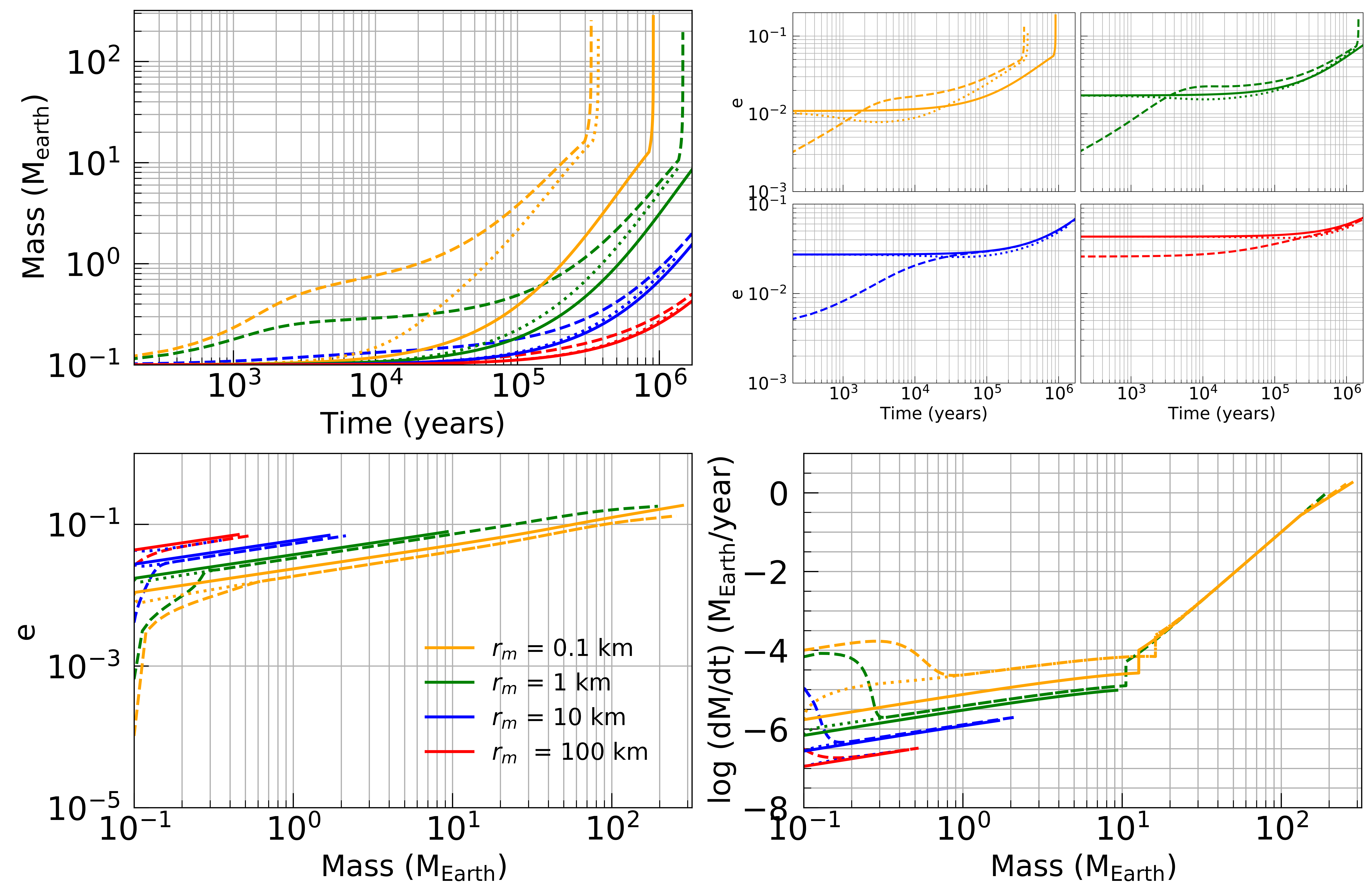}}
\caption{Formation of a planet at 6\,au, without migration. Colors represent planetesimals with different radii (red, 100 km; blue, 10 km; green, 1 km; orange, 0,1 km). The embryo's initial mass is 0.1$\mathrm{M}_{\mathrm{earth}}$ and it is allowed to grow until it reaches 1$\mathrm{M}_{\mathrm{jup}}$ or until time reaches $1.3 \times 10^{6}$ years. The solid line corresponds to the equilibrium scenario (Eq. \ref{eqecceq}). The dashed and dotted lines correspond to the out of equilibrium scenario for two different initial conditions (Eq. \ref{ecctipo2} for the dashed line and Eq. \ref{eqecceq} for the dotted line). The top-left panel shows the planet's mass as a function of time. The top-right panel shows the time evolution of $e$. The bottom-left panel shows the same quantity as a function of the mass of the planet. The bottom-right panel shows the accretion rate, which is the sum of the solids accretion rate and the gas accretion rate, as a function of the planet's mass.}
\label{resumofargoform1MJ}
\end{figure*}

The simulation is interrupted when the planet reaches 1$\mathrm{M}_{\mathrm{jup}}$ or simulation time reaches $1.3 \times 10^{6}$ years, estimated as the remaining lifetime of the disc.  In this work, we do not incorporate photoevaporation of the gas disc, which could change how much gas is available over time. A function to simulate this effect will be developed in future work.

Figure \ref{resumofargoform1MJ} is analogous to Fig. \ref{resform1MT}, but now we allow the planet to grow up to $1\mathrm{M}_{\mathrm{jup}}$ and take gas accretion into account. As done in Section \ref{solidformation}, we use two models for the evolution of eccentricities and inclinations of planetesimals: out of equilibrium scenario (dashed and dotted lines in Fig. \ref{resumofargoform1MJ}) and equilibrium scenario (solid line in Fig. \ref{resumofargoform1MJ}). On all panels, the families of curves are parameterized by the radius of the planetesimals, with their corresponding colors shown in the left-bottom panel.

We note that, for planetesimals with radius $r_{m} = 10$ km and 100 km, the planet grows at a much slower pace (Fig. \ref{resumofargoform1MJ}\,top-left). The final mass, attained after $1.3 \times 10^{6}$ years, is $0.3 \mathrm{M}_{\mathrm{earth}}$ for planetesimals with $r_{m} = 100 $ km radius and $1 \mathrm{M}_{\mathrm{earth}}$ for planetesimals with $r_{m} = 10$ km radius. That is, for these sizes of planetesimals, it is not possible to form a planet with mass of the order of Jupiter's mass (at a fixed position)  during the lifetime of the disc because solids accretion is a long-lasting stage. Since this stage depends strongly on the planetesimals' size, it significantly impacts the planet's final mass.

Conversely, for small planetesimals, with radius of $r_{m} = 0.1 $ km and 1 km, once the critical mass is reached, rapid gas accretion leads to the formation of a giant planet with mass of about one Jupiter mass. Therefore, we conclude that gas drag acting on small planetesimals favours the growth of a solid nucleus up to its critical mass, and that, from this point on, gas accretion is fast enough to form giant planets. In fact, when gas accretion begins, the accretion rate of mass grows abruptly (Fig. \ref{resumofargoform1MJ}, bottom-right). It is worth mentioning that the critical mass depends on the opacity and on the solids accretion rate (see Eq. \ref{formgigantes1}). Thus, variations of these factors due to some additional mechanisms, such as contamination of the planetary envelope, could change its value \citep{Venturini2015}.

The evolution of $e$ (Fig. \ref{resumofargoform1MJ}, top-right) and of the planetary mass (Fig. \ref{resumofargoform1MJ}, bottom-left) as a function of time is similar to that shown  in \cite{Fortier2013}, that is, we observe a similar asymptotic behavior of $e$ as a function of time (Fig. \ref{resumofargoform1MJ}, top-right); we also observe a linear behavior of $e$ as a function of the mass (Fig. \ref{resumofargoform1MJ}, bottom-left). However, there are some differences and one of them is that our model displays an abrupt transition between solids accretion and gas accretion stages, while the model in \cite{Fortier2013} shows a smooth transition because of their equations for the structure of the planetary envelope. It is worth noting that our simplification does not significantly interfere on the final planetary mass, since transition into runaway regime during gas accretion is very rapid. This behaviour has also been detected by
\cite{Ronco2017}, which used a transition to gas accretion similar to ours. 
Thus, although we use a simplified model to transition from solids accretion to gas accretion, our results agree with the literature.

It is worth noting that when the planet's mass is above $10 \mathrm{M}_{\mathrm{earth}}$, gas accretion begins. This decreases gas density around the planet, which in turn, causes and abrupt increase in eccentricity while the core grows. However, there is a physical limitator at which the solids accretion model stops (see Eq. \ref{eqmscat}). So, beyond this limiting mass value, planetesimals are considered to be ejected and do not contribute to the core's formation. So, although the value of $e$ still appears in the plots after the limiting mass value is reached, planetesimals do not longer influence solids accretion.

Similarly to what we observe when a planet with one Earth mass is formed (Fig.~\ref{resform1MT}), in Fig~\ref{resumofargoform1MJ} we also observe that the equilibrium scenario and the out of equilibrium scenario differ more significantly for small planetesimals than for large planetesimals. In fact, for small planetesimals, the out of equilibrium scenario allows to form a planet with 1$\mathrm{M}_{\mathrm{jup}}$ in a shorter time than in the equilibrium scenario. For larger planetesimals, there is no significative difference regarding formation time. Therefore, the equilibrium scenario is a suitable approximation only for large planetesimals.

Thus, our results in this section show that the planetesimals' size in the gas disc has a strong influence on the final mass of the planet, to the point of determining whether a planet becomes a giant planet or not.

\subsection{Migration of planets}\label{mig}

\begin{figure}
\centering
\includegraphics[width=1.0 \columnwidth,angle=0]{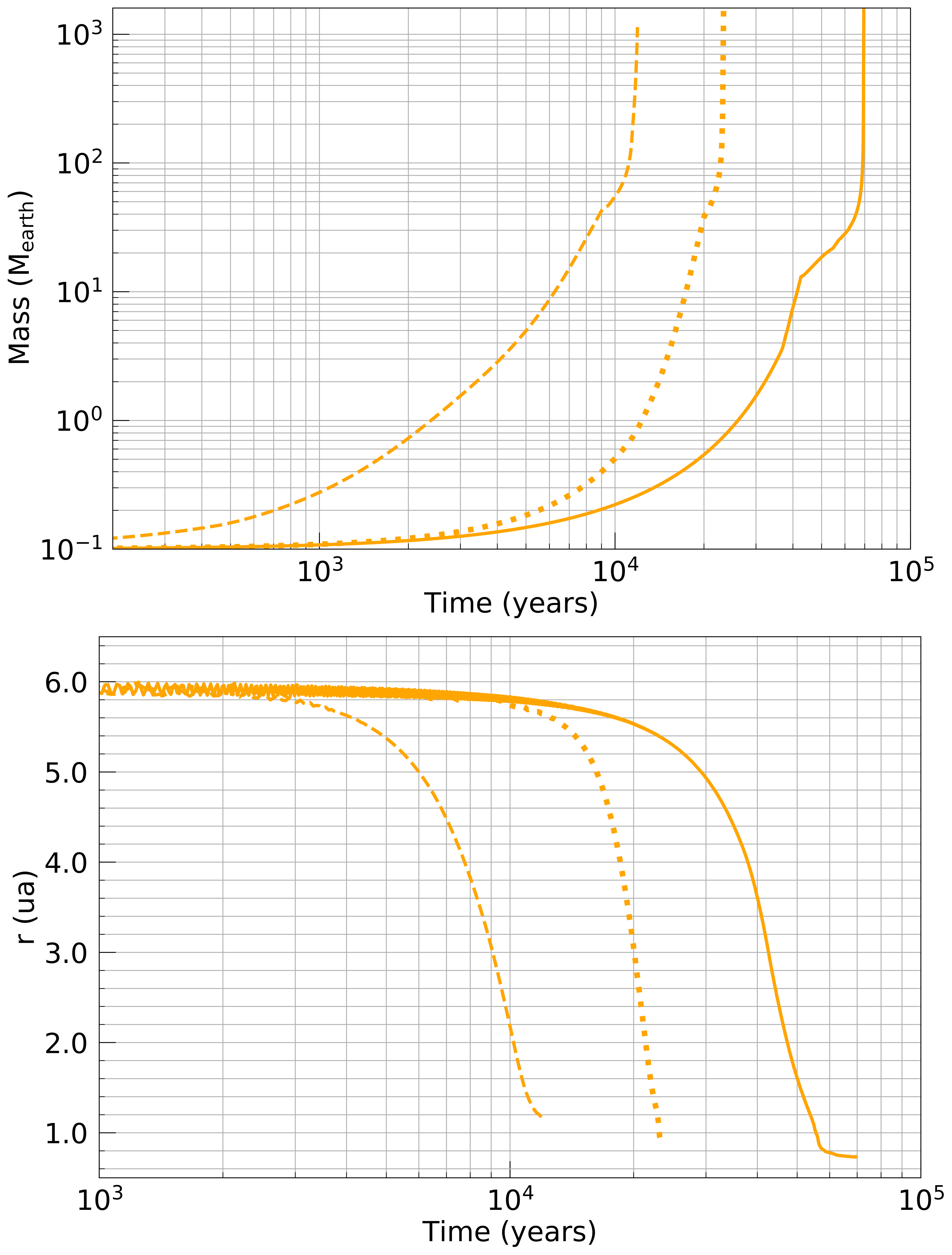}
\caption{Time evolution of the planet's mass (upper panel) and radial position of the planet (bottom panel) of a migrating planet with initial mass of 0.1$\mathrm{M}_{\mathrm{earth}}$ and initial position of 6\,au. The solid line corresponds to the equilibrium scenario (see Eq. \ref{eqecceq}). The dashed and dotted lines correspond to the out of equilibrium scenario for two different initial conditions (see Eq. \ref{ecctipo2} for the dashed line and Eq. \ref{eqecceq} for the dotted line).}
\label{resFARGOFORMMIG}
\end{figure}

In this section we complete our analysis allowing the growing protoplanet to migrate. Figure \ref{resFARGOFORMMIG} shows how the planet's mass (upper panel) and its radial position (bottom panel) evolve with time. The protoplanet's initial mass is 0.1$\mathrm{M}_{\mathrm{earth}}$ and its initial position is 6\,au. The growing protoplanet migrates interacting with a gas disc which contains small planetesimals with radius $r_{m}$ = 0.1 km. This setup follows current models of planetesimal formation,  which indicate that small planetesimals are predominant in protoplanetary discs \citep{Simon2016}. The evolution of the eccentricities of planetesimals is described by the equations corresponding to the out of equilibrium scenario, using two different initial conditions, and also by the equation corresponding to the equilibrium scenario (see Section \ref{evroot}). To avoid disc edge effects and tidal effects due to the host star, the planet is allowed to migrate only up to 0.7\,au. Also, migration is interrupted when a mass equal to 5$\mathrm{M}_{\mathrm{jup}}$ is reached; indeed, once this mass value is attained, the planet enters in type II migration regime, when its position does not change significantly over time. In this work we did not detect type III migration. An investigation of type III migration during the planet formation process is left to future research.

Observing Fig.\,\ref{resFARGOFORMMIG},  in the equilibrium scenario (solid line), we obtain a planet with a mass of approximately 5$\mathrm{M}_{\mathrm{jup}}$ at 0.73\,au, in a time of $6.95 \times 10^4$ years, for planetesimals with $r_{m} = 0.1$ km. Moreover, in the out of equilibrium scenario and a hot disc (dashed line), the planet reaches a mass of 5$\mathrm{M}_{\mathrm{jup}}$ at 1.1\,au, in a time of $ 1.2 \times 10^{4}$ years. For the second initial condition of the out of equilibrium scenario, which corresponds to a cold disc (dotted line), the planet reaches a mass of 5$\mathrm{M}_{\mathrm{jup}}$ at 0.9\,au, in a time of about $2.3 \times 10^4$ years. That is, in all three cases, we obtain giant planets close to the star.

\begin{figure}
\centering
\includegraphics[width=1.0 \columnwidth,angle=0]{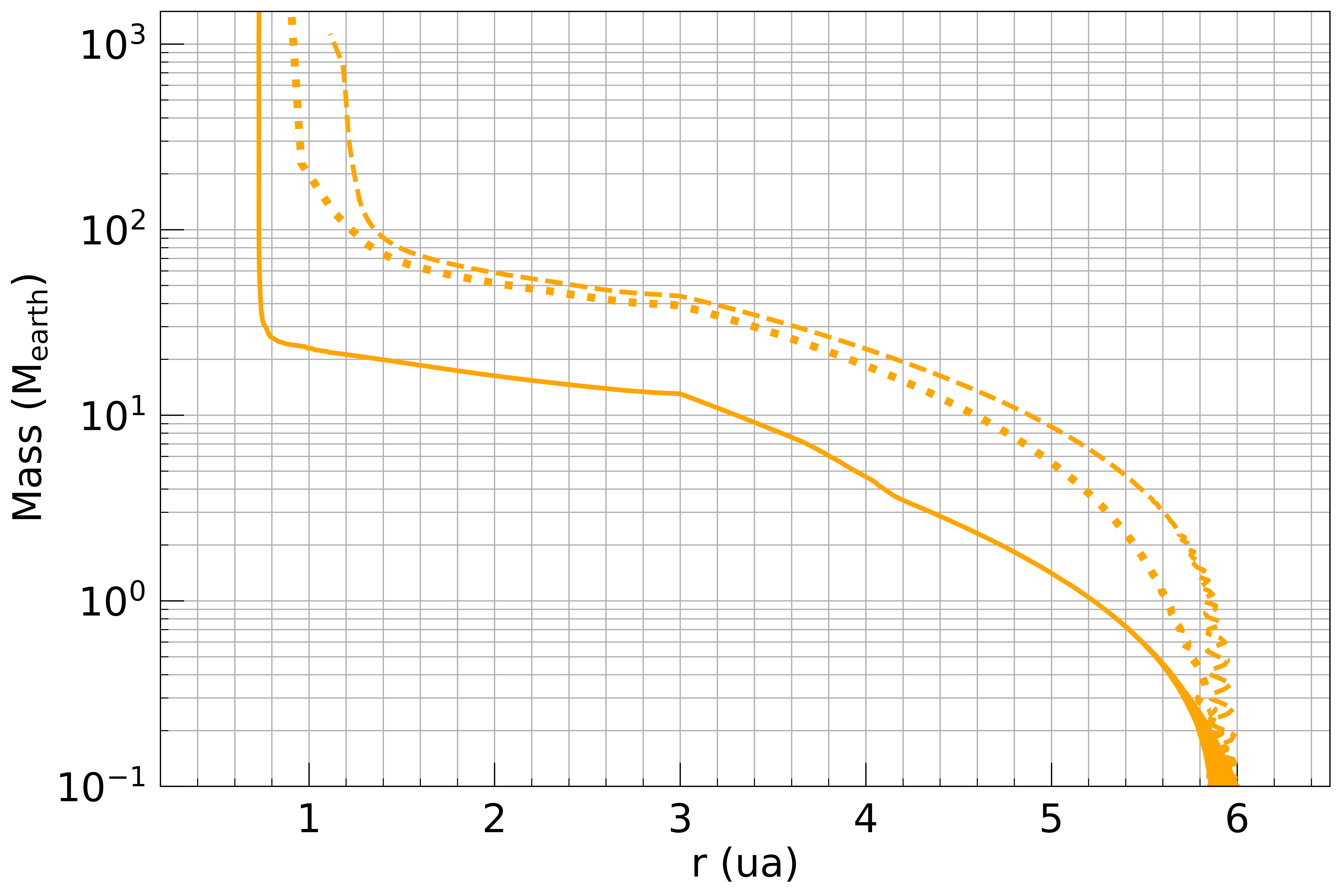}
\caption{Mass of the migrating planet as a function of its position: equilibrium scenario (solid line), out of equilibrium scenario and a hot disc (dashed line), and out of equilibrium scenario and a cold disc (dotted line). The planet starts migrating with mass of 0.1$\mathrm{M}_{\mathrm{earth}}$ at 6\,au. }
\label{massxr}
\end{figure}

Figure~\ref{massxr} shows the mass of the migrating planet as a function of its position for all three situations described above. We observe that the curves present similar behavior in all cases. Contrarily to \cite{Fortier2013}, which uses an adiabatic disc, we do not observe reverse migration. This is because our model uses an isothermal disc. For an adiabatic disc, type I migration could reach a saturated regime, in which the corotation torque could contribute differently, altering the migration rate \citep{Dittkrist2014}.

In Fig.\,\ref{massxr} we note that, when the planet crosses the ice line at 3\,au, the condensation factor ($f_{\mathrm{R/I}}$) changes, decreasing the amount of material available for solids accretion and leading to a noticeable drop in the planet's growth. Despite this fact, the solid core, which is close to its critical value, continues to grow and migrates until gas accretion begins. When the runaway regime for gas accretion begins, planetary mass increases exponentially and the planet enters in type II migration regime. We observe that, in the equilibrium scenario (solid line), the critical mass is about 20$\mathrm{M}_{\mathrm{earth}}$, while, in the out of equilibrium scenario, the critical mass doubles its value up to $\simeq40\mathrm{M}_{\mathrm{earth}}$, for both hot and cold disc (dashed and dotted lines). This results are in agreement with the values obtained in the literature, which vary between 10 and 50$\mathrm{M}_{\mathrm{earth}}$ \citep{Ikoma2000}. However, it is worth noting that the opacity of the envelope could significantly alter the critical mass and the onset of gas capture. We expect to explore these effects in future work.

\begin{figure}
\centering
\includegraphics[width=1.0 \columnwidth,angle=0]{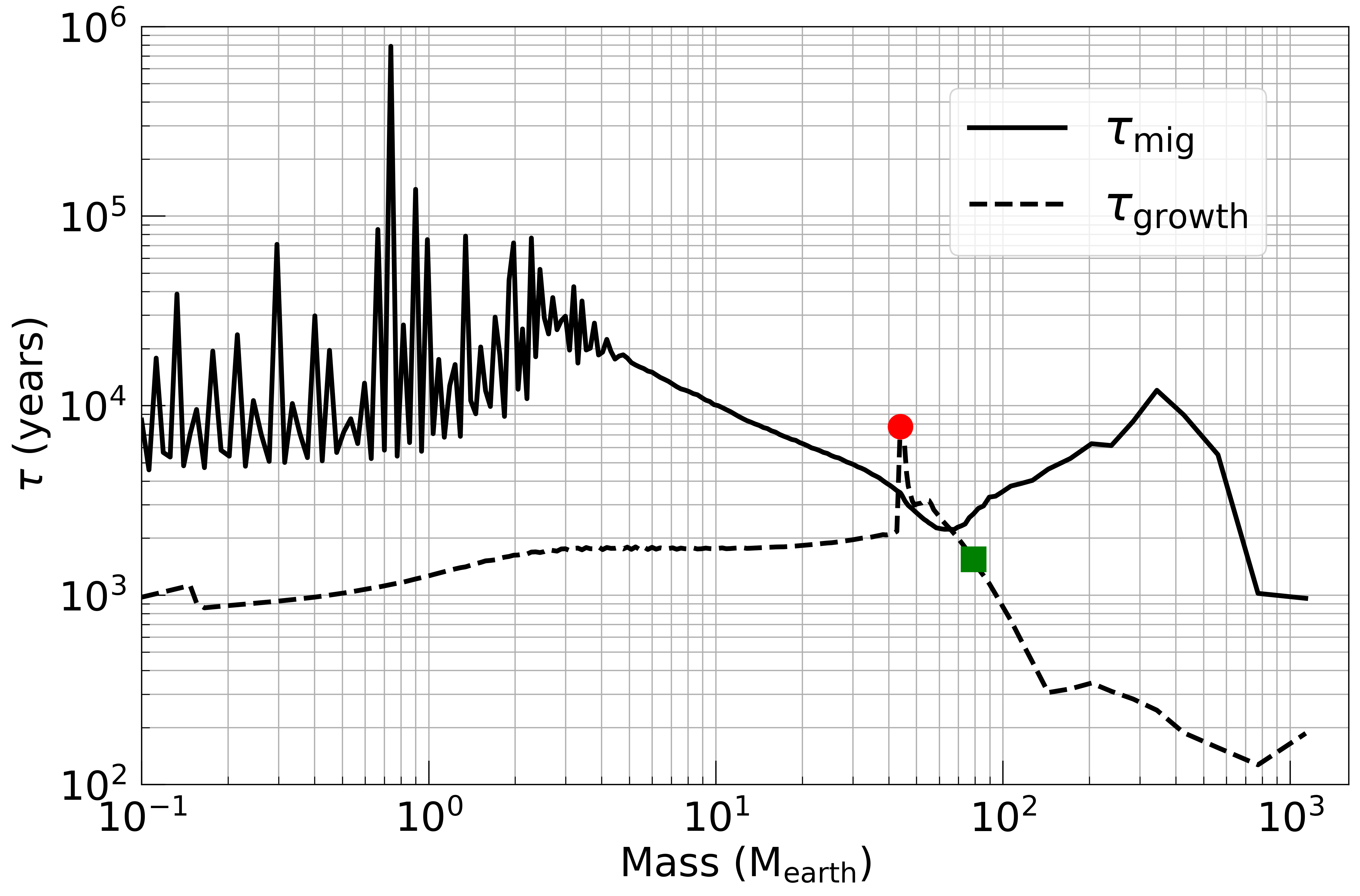}
\caption{Timescale of the planet's growth ($\tau_{\mathrm{growth}} = M/dM_{\mathrm{core}}$; solid line) and timescale of migration ($\tau_{\mathrm{mig}} = a/\dot{a}$; dotted line), for the out of equilibrium scenario in a hot disc. The red dot (to the left) marks the moment at which the planet crosses the ice line, while the green square (to the right) marks when gas accretion begins in runaway regime.}
\label{escalasdetempo}
\end{figure}

Figure \ref{escalasdetempo} allows us to relate the growth of the planet and its migration by comparing the timescales of growth ($\tau_{\mathrm{growth}} = M/dM_{\mathrm{core}}$) and of migration ($\tau_{\mathrm{mig}} = a/\dot{a}$) for the out of equilibrium scenario in a hot disc (dashed line in Fig. \ref{resFARGOFORMMIG}). 
It is worth noticing that, when simulation begins, the planet is very small and has little effect on the gas disc, but as its mass grows it causes a local destabilization in the gas disc. This effect implies in a small oscillation of the planet's position around 6.0 ua (see bottom frame of Fig. \ref{resFARGOFORMMIG}). 
When we compute the migration rate $\tau_{\mathrm{mig}}$, a rapid fluctuation of this measurement is observed due to sudden changes in $\dot{a}$. This effect, which does not affect the global evolution of the planet, can be reduced by considering a less massive gas disc and a smaller initial mass for the planet.

We see that the timescale of migration (solid line) is larger than the timescale of growth (dotted line) until a mass of $\sim$40$\mathrm{M}_{\mathrm{earth}}$ is reached. At this point, both timescales become similar. This is because, until the growing core attains the critical mass,  migration rate increases with the increasing mass, while the timescale of growth increases slowly. When the planet crosses the ice line zone, at 3\,ua, we observe some discontinuity in the smooth evolution of the growth timescale caused by an abrupt change in solid density inside the feeding area. This point is marked in Fig. \ref{escalasdetempo} by a red dot. When the planet crosses the ice line, its core reaches the critical mass and gas accretion begins, the growth timescale becomes smaller than the migration timescale; that is, the planet quickly gains mass and enters type II migration. The point at which the planet initiates gas accretion in runaway regime is marked by a green square in Fig. \ref{escalasdetempo}.

Since the condensation factor affects how much material is available for solids accretion, the ice line has a strong effect in the scenarios described above. Indeed, to avoid that a planet falls into the stellar envelope, the mass of the solid nucleus must be close to its critical value when crossing the ice line zone such that the nucleus will have sufficient time to reach its critical mass and initiate gas accretion, allowing rapid mass growth and entry into type II migration. 

\section{Conclusions}

In this work, we develop a module for planetary formation within the hydrodynamic simulator FARGO3D, using a solid accretion model based on  \cite{Guilera2010,Fortier2013}. 
To exemplify how the module for planetary formation can be used, we analize the effect of the position of the ice line during the formation and migration of a planet. The example given shows that, in order for the planet to enter into type II migration and to avoid falling into the stellar envelope, the mass of the solid nucleus must be close to its critical mass when crossing the ice line.

Adopting an isothermal gas disc, we simulate the formation of terrestrial planets with initial mass of $0.01 \mathrm{M}_{\mathrm{earth}}$ fixed at  6\,au. The results obtained show a strong dependence on the size of planetesimals: planets grow faster when immersed in regions of smaller planetesimals. This occurs because gas drag rapidly decreases the root mean square eccentricities and inclinations of small planetesimals in the feeding zone, allowing the planet to enter into a runaway solid accretion regime faster than in the case of larger planetesimals. Indeed, the latter suffer weaker effects from the gas drag, and maintain their high eccentricities and inclinations for longer time.

We adopt the gas accretion model described in \cite{DePaula2018} and, additionally, we calculate estimates for the critical mass and for the capture radius of planetesimals. It is worth noting that, for this task, the complete gas accretion model would require to solve a system of equations describing the envelope, leading to higher computational costs. However, as shown here, our simple alternative model allows to obtain results which are very similar to the ones obtained using more refined models \citep{Fortier2013}.

To exemplify an application of the model, we study, simultaneously, planetary formation and migration. In particular, we analyze the timescale involved in the migration process in conjunction with the timescale for planetary formation. This example reveals that, for the set of parameters chosen in this work, it is possible to obtain a planet's growth timescale shorter than the timescale of migration, even when the planet crosses the ice line during its inward migration.  It also shows that, for planetesimals with radius $\sim$0.1\,km, it is possible to obtain planets that grow up to approximately 5 times the mass of Jupiter in regions between 0.5\,au and 1\,au in times shorter than the estimated lifetime of the gas disc ($\sim$10\,Myr).

Besides that, we note that the size of planetesimals has an important influence on how the semi-axis and the mass of the planet evolve. This is because their size impacts strongly on the solids accretion rate and, consequently, on the relationship between the timescales of formation and migration, which are strongly coupled.

In order to make our model more realistic, some adaptations  could be implemented immediately into the model presented here. The first one is a variable size for the feeding zone. Here, we use a feeding zone that depends on the Hill's radius of the planet multiplied by a constant factor ($ b = 10 $). However, the feeding zone could depend on the eccentricities and inclinations of the planetesimals. The second characteristic that could be adjusted in the model, is a differential distribution of planetesimals, instead of working with planetesimals of fixed sizes. A differential size distribution could be important to describe how planetary systems are formed as discussed in  \citep{Guilera2010}.

One important point that should yet be explored is the evolution of the ice line. In fact, as shown in Section \ref{mig}, the ice line has a strong effect in planetary formation with migration. In order for a planet not to fall into the stellar envelope, the mass of the solid nucleus must be close to its critical value when crossing the ice line, so there is enough time for it to initiate gas accretion, allowing rapid mass growth and entry into type II migration.

Moreover, the modeling of an adiabatic disc instead of an isothermal disc, including heating in the planet's local region due to planetary accretion, could bring new interesting results. Heating could reduce the rate of type I migration and even reverse migration, as shown in \cite{Benitez2015}. In addition, local heating could be used to obtain radiative feedback, which can be exploited to obtain observational characteristics \citep{Montesinos2015}.


\section*{Acknowledgements}
The authors are grateful to P. Ben\'itez-Llambay for meaningful discussions and suggestions. Also, the authors aknowledge the contribution of Octavio M. Guilera who kindly verified some of the early results and provided important suggestions that lead to
the development of the present work, especially regarding the computation of the capture radius. The authors also thank the reviewer for detailed comments and suggestions that helped improve this manuscript. This work was supported by the Brazilian agencies Conselho Nacional de Desenvolvimento Cient\'ifico e Tecnol\'ogico (CNPq),  and the S\~ao Paulo Research Foundation (FAPESP) (grants  2014/00492-3 and 2016/13750-6). This work has used the computing facilities of the Laboratory of Astroinformatics (IAG/USP, NAT/Unicsul), that were purchased thanks to the Brazilian agency FAPESP (grant 2009/54006-4) and the INCT-A.




\bibliographystyle{mnras}
\bibliography{bibliografia}

\begin{thebibliography}{}
\makeatletter
\relax
\def\mn@urlcharsother{\let\do\@makeother \do\$\do\&\do\#\do\^\do\_\do\%\do\~}
\def\mn@doi{\begingroup\mn@urlcharsother \@ifnextchar [ {\mn@doi@}
  {\mn@doi@[]}}
\def\mn@doi@[#1]#2{\def\@tempa{#1}\ifx\@tempa\@empty \href
  {http://dx.doi.org/#2} {doi:#2}\else \href {http://dx.doi.org/#2} {#1}\fi
  \endgroup}
\def\mn@eprint#1#2{\mn@eprint@#1:#2::\@nil}
\def\mn@eprint@arXiv#1{\href {http://arxiv.org/abs/#1} {{\tt arXiv:#1}}}
\def\mn@eprint@dblp#1{\href {http://dblp.uni-trier.de/rec/bibtex/#1.xml}
  {dblp:#1}}
\def\mn@eprint@#1:#2:#3:#4\@nil{\def\@tempa {#1}\def\@tempb {#2}\def\@tempc
  {#3}\ifx \@tempc \@empty \let \@tempc \@tempb \let \@tempb \@tempa \fi \ifx
  \@tempb \@empty \def\@tempb {arXiv}\fi \@ifundefined
  {mn@eprint@\@tempb}{\@tempb:\@tempc}{\expandafter \expandafter \csname
  mn@eprint@\@tempb\endcsname \expandafter{\@tempc}}}

\bibitem[\protect\citeauthoryear{{Alibert}, {Mordasini}, {Benz}  \&
  {Winisdoerffer}}{{Alibert} et~al.}{2005}]{Alibert2005}
{Alibert} Y.,  {Mordasini} C.,  {Benz} W.,   {Winisdoerffer} C.,  2005, \mn@doi
  [\aap] {10.1051/0004-6361:20042032}, \href
  {http://adsabs.harvard.edu/abs/2005A%26A...434..343A} {434, 343}

\bibitem[\protect\citeauthoryear{{Armitage}}{{Armitage}}{2010}]{Armitage2010}
{Armitage} P.~J.,  2010, {Astrophysics of Planet Formation}

\bibitem[\protect\citeauthoryear{{Baruteau} et~al.,}{{Baruteau}
  et~al.}{2014}]{Baruteau2014}
{Baruteau} C.,  et~al., 2014, \mn@doi [Protostars and Planets VI]
  {10.2458/azu_uapress_9780816531240-ch029}, \href
  {http://adsabs.harvard.edu/abs/2014prpl.conf..667B} {pp 667--689}

\bibitem[\protect\citeauthoryear{{Ben{\'{\i}}tez-Llambay} \&
  {Masset}}{{Ben{\'{\i}}tez-Llambay} \& {Masset}}{2016}]{Benitez2016}
{Ben{\'{\i}}tez-Llambay} P.,  {Masset} F.~S.,  2016, \mn@doi [\apjs]
  {10.3847/0067-0049/223/1/11}, \href
  {http://adsabs.harvard.edu/abs/2016ApJS..223...11B} {223, 11}

\bibitem[\protect\citeauthoryear{{Ben{\'{\i}}tez-Llambay}, {Masset},
  {Koenigsberger}  \& {Szul{\'a}gyi}}{{Ben{\'{\i}}tez-Llambay}
  et~al.}{2015}]{Benitez2015}
{Ben{\'{\i}}tez-Llambay} P.,  {Masset} F.,  {Koenigsberger} G.,
  {Szul{\'a}gyi} J.,  2015, \mn@doi [\nat] {10.1038/nature14277}, \href
  {http://adsabs.harvard.edu/abs/2015Natur.520...63B} {520, 63}

\bibitem[\protect\citeauthoryear{{Bitsch} \& {Kley}}{{Bitsch} \&
  {Kley}}{2010}]{Bitsch2010}
{Bitsch} B.,  {Kley} W.,  2010, \mn@doi [\aap] {10.1051/0004-6361/201014414},
  \href {http://adsabs.harvard.edu/abs/2010A%26A...523A..30B} {523, A30}

\bibitem[\protect\citeauthoryear{{Bryden} \& {Lin}}{{Bryden} \&
  {Lin}}{1999}]{Bryden1999}
{Bryden} G.,  {Lin} D.~N.~C.,  1999, in Bulletin of the American Astronomical
  Society. p.~1130

\bibitem[\protect\citeauthoryear{{Chambers}}{{Chambers}}{2006}]{Chambers2006}
{Chambers} J.,  2006, \mn@doi [\icarus] {10.1016/j.icarus.2005.10.017}, \href
  {http://adsabs.harvard.edu/abs/2006Icar..180..496C} {180, 496}

\bibitem[\protect\citeauthoryear{{Crida}, {Baruteau}, {Kley}  \&
  {Masset}}{{Crida} et~al.}{2009}]{Crida2009}
{Crida} A.,  {Baruteau} C.,  {Kley} W.,   {Masset} F.,  2009, \mn@doi [\aap]
  {10.1051/0004-6361/200811608}, \href
  {http://adsabs.harvard.edu/abs/2009A%26A...502..679C} {502, 679}

\bibitem[\protect\citeauthoryear{{D'Angelo} \& {Lubow}}{{D'Angelo} \&
  {Lubow}}{2010}]{Angelo2010}
{D'Angelo} G.,  {Lubow} S.~H.,  2010, \mn@doi [\apj]
  {10.1088/0004-637X/724/1/730}, \href
  {https://ui.adsabs.harvard.edu/abs/2010ApJ...724..730D} {724, 730}

\bibitem[\protect\citeauthoryear{DePaula \& Michtchenko}{DePaula \&
  Michtchenko}{2018}]{DePaula2018}
DePaula L.~A.,  Michtchenko T.~A.,  2018, Monthly Notices of the Royal
  Astronomical Society, 483, 1599

\bibitem[\protect\citeauthoryear{{Dittkrist}, {Mordasini}, {Klahr}, {Alibert}
  \& {Henning}}{{Dittkrist} et~al.}{2014}]{Dittkrist2014}
{Dittkrist} K.-M.,  {Mordasini} C.,  {Klahr} H.,  {Alibert} Y.,   {Henning} T.,
   2014, \mn@doi [\aap] {10.1051/0004-6361/201322506}, \href
  {http://cdsads.u-strasbg.fr/abs/2014A%26A...567A.121D} {567, A121}

\bibitem[\protect\citeauthoryear{{D{\"u}rmann} \& {Kley}}{{D{\"u}rmann} \&
  {Kley}}{2015}]{Durman2015}
{D{\"u}rmann} C.,  {Kley} W.,  2015, \mn@doi [\aap]
  {10.1051/0004-6361/201424837}, \href
  {http://adsabs.harvard.edu/abs/2015A%26A...574A..52D} {574, A52}

\bibitem[\protect\citeauthoryear{{Fortier}, {Alibert}, {Carron}, {Benz}  \&
  {Dittkrist}}{{Fortier} et~al.}{2013}]{Fortier2013}
{Fortier} A.,  {Alibert} Y.,  {Carron} F.,  {Benz} W.,   {Dittkrist} K.-M.,
  2013, \mn@doi [\aap] {10.1051/0004-6361/201220241}, \href
  {http://adsabs.harvard.edu/abs/2013A%26A...549A..44F} {549, A44}

\bibitem[\protect\citeauthoryear{{Goldreich} \& {Tremaine}}{{Goldreich} \&
  {Tremaine}}{1980}]{Goldreich1980}
{Goldreich} P.,  {Tremaine} S.,  1980, \mn@doi [\apj] {10.1086/158356}, \href
  {http://adsabs.harvard.edu/abs/1980ApJ...241..425G} {241, 425}

\bibitem[\protect\citeauthoryear{{Guilera}}{{Guilera}}{2016}]{Guilera2016}
{Guilera} O.~M.,  2016, Boletin de la Asociacion Argentina de Astronomia La
  Plata Argentina, \href {http://adsabs.harvard.edu/abs/2016BAAA...58..316G}
  {58, 316}

\bibitem[\protect\citeauthoryear{{Guilera}, {Brunini}  \&
  {Benvenuto}}{{Guilera} et~al.}{2010}]{Guilera2010}
{Guilera} O.~M.,  {Brunini} A.,   {Benvenuto} O.~G.,  2010, \mn@doi [\aap]
  {10.1051/0004-6361/201014365}, \href
  {http://adsabs.harvard.edu/abs/2010A%26A...521A..50G} {521, A50}

\bibitem[\protect\citeauthoryear{{Ida} \& {Lin}}{{Ida} \&
  {Lin}}{2004a}]{Ida2004a}
{Ida} S.,  {Lin} D.~N.~C.,  2004a, \mn@doi [\apj] {10.1086/381724}, \href
  {http://adsabs.harvard.edu/abs/2004ApJ...604..388I} {604, 388}

\bibitem[\protect\citeauthoryear{{Ida} \& {Lin}}{{Ida} \&
  {Lin}}{2004b}]{Ida2004b}
{Ida} S.,  {Lin} D.~N.~C.,  2004b, \mn@doi [\apj] {10.1086/424830}, \href
  {http://adsabs.harvard.edu/abs/2004ApJ...616..567I} {616, 567}

\bibitem[\protect\citeauthoryear{{Ida} \& {Lin}}{{Ida} \&
  {Lin}}{2008}]{Ida2008}
{Ida} S.,  {Lin} D.~N.~C.,  2008, \mn@doi [\apj] {10.1086/523754}, \href
  {http://adsabs.harvard.edu/abs/2008ApJ...673..487I} {673, 487}

\bibitem[\protect\citeauthoryear{{Ikoma}, {Nakazawa}  \& {Emori}}{{Ikoma}
  et~al.}{2000}]{Ikoma2000}
{Ikoma} M.,  {Nakazawa} K.,   {Emori} H.,  2000, \mn@doi [\apj]
  {10.1086/309050}, \href {http://adsabs.harvard.edu/abs/2000ApJ...537.1013I}
  {537, 1013}

\bibitem[\protect\citeauthoryear{{Inaba}, {Tanaka}, {Nakazawa}, {Wetherill}  \&
  {Kokubo}}{{Inaba} et~al.}{2001}]{Inaba2001}
{Inaba} S.,  {Tanaka} H.,  {Nakazawa} K.,  {Wetherill} G.~W.,   {Kokubo} E.,
  2001, \mn@doi [\icarus] {10.1006/icar.2000.6533}, \href
  {http://adsabs.harvard.edu/abs/2001Icar..149..235I} {149, 235}

\bibitem[\protect\citeauthoryear{{Johansen} \& {Lambrechts}}{{Johansen} \&
  {Lambrechts}}{2017}]{Johansen2017}
{Johansen} A.,  {Lambrechts} M.,  2017, \mn@doi [Annual Review of Earth and
  Planetary Sciences] {10.1146/annurev-earth-063016-020226}, \href
  {http://adsabs.harvard.edu/abs/2017AREPS..45..359J} {45, 359}

\bibitem[\protect\citeauthoryear{{Kley}}{{Kley}}{1999}]{Kley1999}
{Kley} W.,  1999, \mn@doi [\mnras] {10.1046/j.1365-8711.1999.02198.x}, \href
  {http://adsabs.harvard.edu/abs/1999MNRAS.303..696K} {303, 696}

\bibitem[\protect\citeauthoryear{{Kley} \& {Nelson}}{{Kley} \&
  {Nelson}}{2012}]{Kley2012}
{Kley} W.,  {Nelson} R.~P.,  2012, \mn@doi [\araa]
  {10.1146/annurev-astro-081811-125523}, \href
  {http://adsabs.harvard.edu/abs/2012ARA%26A..50..211K} {50, 211}

\bibitem[\protect\citeauthoryear{{Lambrechts} \& {Johansen}}{{Lambrechts} \&
  {Johansen}}{2014}]{Lambrechts2014}
{Lambrechts} M.,  {Johansen} A.,  2014, \mn@doi [\aap]
  {10.1051/0004-6361/201424343}, \href
  {http://adsabs.harvard.edu/abs/2014A%26A...572A.107L} {572, A107}

\bibitem[\protect\citeauthoryear{{Lin} \& {Papaloizou}}{{Lin} \&
  {Papaloizou}}{2010}]{Lin2010}
{Lin} M.-K.,  {Papaloizou} J.~C.~B.,  2010, \mn@doi [\mnras]
  {10.1111/j.1365-2966.2010.16560.x}, \href
  {https://ui.adsabs.harvard.edu/abs/2010MNRAS.405.1473L} {405, 1473}

\bibitem[\protect\citeauthoryear{{Lissauer}}{{Lissauer}}{1993}]{Lissauer1993}
{Lissauer} J.~J.,  1993, \mn@doi [\araa] {10.1146/annurev.aa.31.090193.001021},
  \href {http://adsabs.harvard.edu/abs/1993ARA%26A..31..129L} {31, 129}

\bibitem[\protect\citeauthoryear{{Martin} \& {Livio}}{{Martin} \&
  {Livio}}{2012}]{Martin2012}
{Martin} R.~G.,  {Livio} M.,  2012, \mn@doi [\mnras]
  {10.1111/j.1745-3933.2012.01290.x}, \href
  {https://ui.adsabs.harvard.edu/abs/2012MNRAS.425L...6M} {425, L6}

\bibitem[\protect\citeauthoryear{{Martin} \& {Livio}}{{Martin} \&
  {Livio}}{2013}]{Martin2013}
{Martin} R.~G.,  {Livio} M.,  2013, \mn@doi [\mnras] {10.1093/mnras/stt1051},
  \href {https://ui.adsabs.harvard.edu/abs/2013MNRAS.434..633M} {434, 633}

\bibitem[\protect\citeauthoryear{{Masset}}{{Masset}}{2000}]{Masset2000A}
{Masset} F.,  2000, \mn@doi [\aaps] {10.1051/aas:2000116}, \href
  {http://adsabs.harvard.edu/abs/2000A%26AS..141..165M} {141, 165}

\bibitem[\protect\citeauthoryear{{Meyer-Vernet} \& {Sicardy}}{{Meyer-Vernet} \&
  {Sicardy}}{1987}]{Meyer1987}
{Meyer-Vernet} N.,  {Sicardy} B.,  1987, \mn@doi [\icarus]
  {10.1016/0019-1035(87)90011-X}, \href
  {http://adsabs.harvard.edu/abs/1987Icar...69..157M} {69, 157}

\bibitem[\protect\citeauthoryear{{Mignone}, {Zanni}, {Tzeferacos}, {van
  Straalen}, {Colella}  \& {Bodo}}{{Mignone} et~al.}{2012}]{Mignone2012}
{Mignone} A.,  {Zanni} C.,  {Tzeferacos} P.,  {van Straalen} B.,  {Colella} P.,
    {Bodo} G.,  2012, \mn@doi [\apjs] {10.1088/0067-0049/198/1/7}, \href
  {http://adsabs.harvard.edu/abs/2012ApJS..198....7M} {198, 7}

\bibitem[\protect\citeauthoryear{{Min}, {Dullemond}, {Kama}  \&
  {Dominik}}{{Min} et~al.}{2011}]{Min2011}
{Min} M.,  {Dullemond} C.~P.,  {Kama} M.,   {Dominik} C.,  2011, \mn@doi
  [\icarus] {10.1016/j.icarus.2010.12.002}, \href
  {http://adsabs.harvard.edu/abs/2011Icar..212..416M} {212, 416}

\bibitem[\protect\citeauthoryear{{Mizuno}}{{Mizuno}}{1980}]{Mizuno1980}
{Mizuno} H.,  1980, \mn@doi [Progress of Theoretical Physics]
  {10.1143/PTP.64.544}, \href
  {http://adsabs.harvard.edu/abs/1980PThPh..64..544M} {64, 544}

\bibitem[\protect\citeauthoryear{{Montesinos}, {Cuadra}, {Perez}, {Baruteau}
  \& {Casassus}}{{Montesinos} et~al.}{2015}]{Montesinos2015}
{Montesinos} M.,  {Cuadra} J.,  {Perez} S.,  {Baruteau} C.,   {Casassus} S.,
  2015, \mn@doi [\apj] {10.1088/0004-637X/806/2/253}, \href
  {http://adsabs.harvard.edu/abs/2015ApJ...806..253M} {806, 253}

\bibitem[\protect\citeauthoryear{{Mordasini}, {Klahr}, {Alibert}, {Benz}  \&
  {Dittkrist}}{{Mordasini} et~al.}{2010}]{Mordasini2010}
{Mordasini} C.,  {Klahr} H.,  {Alibert} Y.,  {Benz} W.,   {Dittkrist} K.-M.,
  2010, preprint, \href {http://adsabs.harvard.edu/abs/2010arXiv1012.5281M} {}
  (\mn@eprint {arXiv} {1012.5281})

\bibitem[\protect\citeauthoryear{{Movshovitz} \& {Podolak}}{{Movshovitz} \&
  {Podolak}}{2008}]{Movshovitz2008}
{Movshovitz} N.,  {Podolak} M.,  2008, \mn@doi [\icarus]
  {10.1016/j.icarus.2007.09.018}, \href
  {https://ui.adsabs.harvard.edu/abs/2008Icar..194..368M} {194, 368}

\bibitem[\protect\citeauthoryear{{Ohtsuki}, {Stewart}  \& {Ida}}{{Ohtsuki}
  et~al.}{2002}]{Ohtsuki2002}
{Ohtsuki} K.,  {Stewart} G.~R.,   {Ida} S.,  2002, \mn@doi [\icarus]
  {10.1006/icar.2001.6741}, \href
  {http://adsabs.harvard.edu/abs/2002Icar..155..436O} {155, 436}

\bibitem[\protect\citeauthoryear{{Ormel} \& {Kobayashi}}{{Ormel} \&
  {Kobayashi}}{2012}]{Ormel2012}
{Ormel} C.~W.,  {Kobayashi} H.,  2012, \mn@doi [\apj]
  {10.1088/0004-637X/747/2/115}, \href
  {http://adsabs.harvard.edu/abs/2012ApJ...747..115O} {747, 115}

\bibitem[\protect\citeauthoryear{{Paardekooper}, {Baruteau}, {Crida}  \&
  {Kley}}{{Paardekooper} et~al.}{2010}]{Paardekooper2010}
{Paardekooper} S.-J.,  {Baruteau} C.,  {Crida} A.,   {Kley} W.,  2010, \mn@doi
  [\mnras] {10.1111/j.1365-2966.2009.15782.x}, \href
  {http://adsabs.harvard.edu/abs/2010MNRAS.401.1950P} {401, 1950}

\bibitem[\protect\citeauthoryear{{Paardekooper}, {Baruteau}  \&
  {Kley}}{{Paardekooper} et~al.}{2011}]{Paardekooper2011}
{Paardekooper} S.-J.,  {Baruteau} C.,   {Kley} W.,  2011, \mn@doi [\mnras]
  {10.1111/j.1365-2966.2010.17442.x}, \href
  {http://adsabs.harvard.edu/abs/2011MNRAS.410..293P} {410, 293}

\bibitem[\protect\citeauthoryear{{Papaloizou}, {Nelson}, {Kley}, {Masset}  \&
  {Artymowicz}}{{Papaloizou} et~al.}{2007}]{Papaloizou2007}
{Papaloizou} J.~C.~B.,  {Nelson} R.~P.,  {Kley} W.,  {Masset} F.~S.,
  {Artymowicz} P.,  2007, Protostars and Planets V, \href
  {http://adsabs.harvard.edu/abs/2007prpl.conf..655P} {pp 655--668}

\bibitem[\protect\citeauthoryear{{Pepli{\'n}ski}, {Artymowicz}  \&
  {Mellema}}{{Pepli{\'n}ski} et~al.}{2008}]{Peplinski2008}
{Pepli{\'n}ski} A.,  {Artymowicz} P.,   {Mellema} G.,  2008, \mn@doi [\mnras]
  {10.1111/j.1365-2966.2008.13045.x}, \href
  {https://ui.adsabs.harvard.edu/abs/2008MNRAS.386..164P} {386, 164}

\bibitem[\protect\citeauthoryear{{Pollack}, {Hubickyj}, {Bodenheimer},
  {Lissauer}, {Podolak}  \& {Greenzweig}}{{Pollack} et~al.}{1996}]{Pollack1996}
{Pollack} J.~B.,  {Hubickyj} O.,  {Bodenheimer} P.,  {Lissauer} J.~J.,
  {Podolak} M.,   {Greenzweig} Y.,  1996, \mn@doi [\icarus]
  {10.1006/icar.1996.0190}, \href
  {http://adsabs.harvard.edu/abs/1996Icar..124...62P} {124, 62}

\bibitem[\protect\citeauthoryear{{Rafikov}}{{Rafikov}}{2004}]{Rafikov2004}
{Rafikov} R.~R.,  2004, \mn@doi [\aj] {10.1086/423216}, \href
  {http://adsabs.harvard.edu/abs/2004AJ....128.1348R} {128, 1348}

\bibitem[\protect\citeauthoryear{{Ronco}, {Guilera}  \& {de
  El{\'{\i}}a}}{{Ronco} et~al.}{2017}]{Ronco2017}
{Ronco} M.~P.,  {Guilera} O.~M.,   {de El{\'{\i}}a} G.~C.,  2017, \mn@doi
  [\mnras] {10.1093/mnras/stx1746}, \href
  {http://adsabs.harvard.edu/abs/2017MNRAS.471.2753R} {471, 2753}

\bibitem[\protect\citeauthoryear{{Shakura} \& {Sunyaev}}{{Shakura} \&
  {Sunyaev}}{1973}]{Shakura1973}
{Shakura} N.~I.,  {Sunyaev} R.~A.,  1973, \aap, \href
  {http://adsabs.harvard.edu/abs/1973A%26A....24..337S} {24, 337}

\bibitem[\protect\citeauthoryear{{Simon}, {Armitage}, {Li}  \&
  {Youdin}}{{Simon} et~al.}{2016}]{Simon2016}
{Simon} J.~B.,  {Armitage} P.~J.,  {Li} R.,   {Youdin} A.~N.,  2016, \mn@doi
  [\apj] {10.3847/0004-637X/822/1/55}, \href
  {http://adsabs.harvard.edu/abs/2016ApJ...822...55S} {822, 55}

\bibitem[\protect\citeauthoryear{{Stone} \& {Norman}}{{Stone} \&
  {Norman}}{1992}]{Stone1992a}
{Stone} J.~M.,  {Norman} M.~L.,  1992, \mn@doi [\apjs] {10.1086/191680}, \href
  {http://adsabs.harvard.edu/abs/1992ApJS...80..753S} {80, 753}

\bibitem[\protect\citeauthoryear{{Takeuchi} \& {Lin}}{{Takeuchi} \&
  {Lin}}{2002}]{Takeuchi2002}
{Takeuchi} T.,  {Lin} D.~N.~C.,  2002, \mn@doi [\apj] {10.1086/344437}, \href
  {http://adsabs.harvard.edu/abs/2002ApJ...581.1344T} {581, 1344}

\bibitem[\protect\citeauthoryear{{Tanaka}, {Takeuchi}  \& {Ward}}{{Tanaka}
  et~al.}{2002}]{Tanaka2002}
{Tanaka} H.,  {Takeuchi} T.,   {Ward} W.~R.,  2002, \mn@doi [\apj]
  {10.1086/324713}, \href {http://adsabs.harvard.edu/abs/2002ApJ...565.1257T}
  {565, 1257}

\bibitem[\protect\citeauthoryear{{Thommes}, {Duncan}  \& {Levison}}{{Thommes}
  et~al.}{2003}]{Thommes2003}
{Thommes} E.~W.,  {Duncan} M.~J.,   {Levison} H.~F.,  2003, \mn@doi [\icarus]
  {10.1016/S0019-1035(02)00043-X}, \href
  {http://adsabs.harvard.edu/abs/2003Icar..161..431T} {161, 431}

\bibitem[\protect\citeauthoryear{{Venturini}, {Alibert}, {Benz}  \&
  {Ikoma}}{{Venturini} et~al.}{2015}]{Venturini2015}
{Venturini} J.,  {Alibert} Y.,  {Benz} W.,   {Ikoma} M.,  2015, \mn@doi [\aap]
  {10.1051/0004-6361/201424008}, \href
  {http://adsabs.harvard.edu/abs/2015A%26A...576A.114V} {576, A114}

\bibitem[\protect\citeauthoryear{{Ward}}{{Ward}}{1986}]{Ward1986}
{Ward} W.~R.,  1986, \mn@doi [\icarus] {10.1016/0019-1035(86)90182-X}, \href
  {http://adsabs.harvard.edu/abs/1986Icar...67..164W} {67, 164}

\bibitem[\protect\citeauthoryear{{Ward}}{{Ward}}{1997}]{Ward1997}
{Ward} W.~R.,  1997, \mn@doi [\icarus] {10.1006/icar.1996.5647}, \href
  {http://adsabs.harvard.edu/abs/1997Icar..126..261W} {126, 261}

\makeatother
\end{thebibliography}




\appendix

\section{Test of resolution for runaway gas accretion}\label{app}

We recommend the reader some caution regarding grid resolution when using the module for planetary formation. It is known that gas accretion models depend on the number of cells within the planet's Hill region. So, it is important to always choose an adequate resolution according to the case under investigation.

\begin{figure}
\centering
\includegraphics[width=1.1 \columnwidth,angle=0]{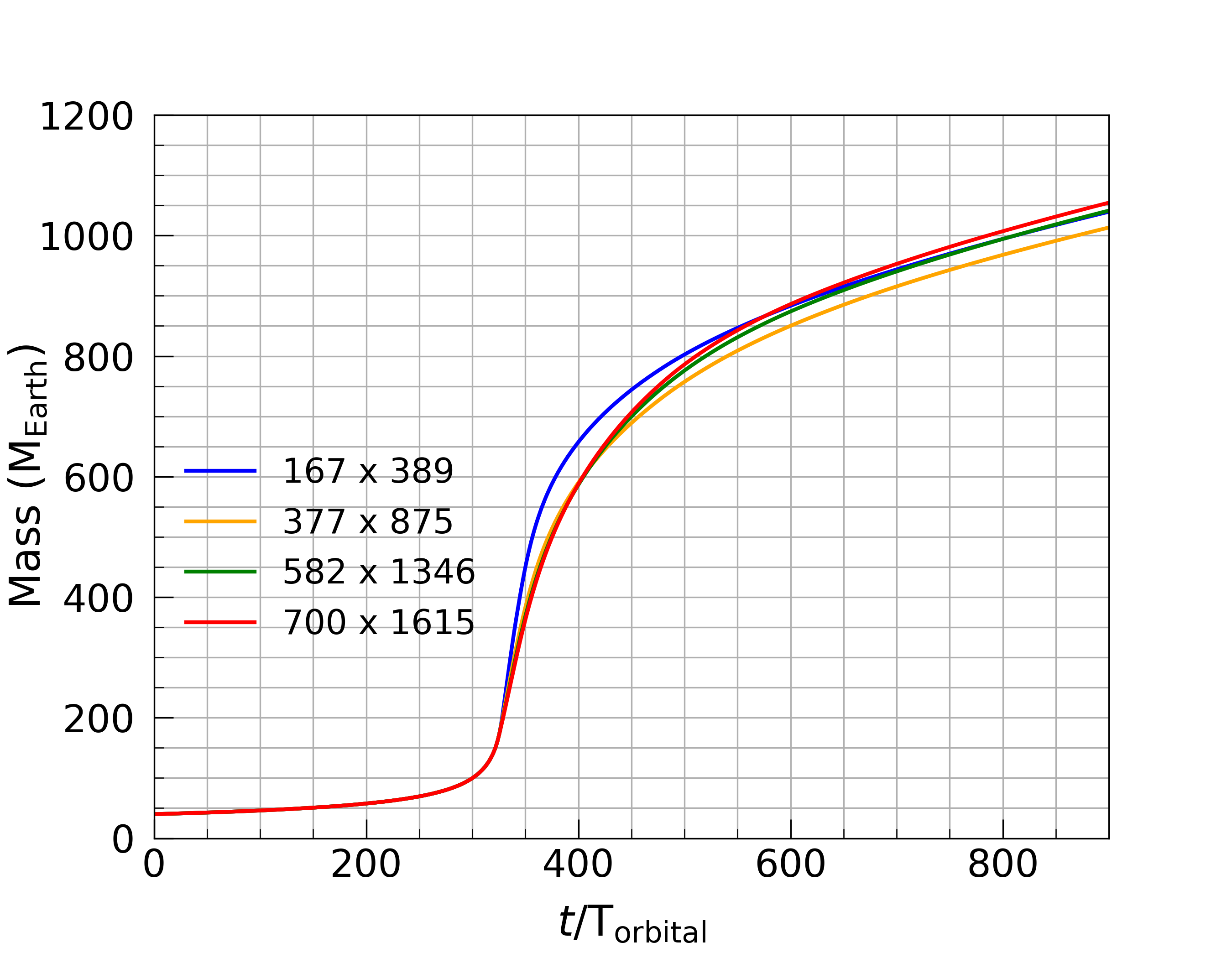}
\caption{Resolution test for the gas accretion model. The planet's initial mass is 30$\mathrm{M}_{\mathrm{earth}}$ and the dimensionless time is defined in terms of the number of orbits of the planet fixed at $r$ = 1 code units (5.2\,au).}
\label{resolution}
\end{figure}

Figure~\ref{resolution} shows a resolution test for the gas accretion model with no planetary migration. That is, we disable the solids accretion model, so that the planet gains mass only through gas accretion. The initial mass of the planet is 30$\mathrm{M}_{\mathrm{earth}}$ and it is fixed at 5.2\,au. For the first resolution (167 $\times$ 389) there are 4 cells inside Hill's region when the simulation begins. This value increases to 12 and 18 for resolutions 377 $\times$ 875 and 582 $\times$ 1346, respectively. Finally, for the last resolution (700 $\times$ 1615), there are 22 cells inside Hill's region when the simulation begins. Analyzing Fig.~\ref{resolution}, we see a good agreement between the final three resolutions up to 650$\mathrm{M}_{\mathrm{earth}}$ ($\backsimeq2\mathrm{M}_{\mathrm{jup}}$). However, as the planet's mass grows, we note that a higher resolution is required for the results to converge.

In this work, we use an equally spaced hydrodynamic grid with resolution of 582 $\times$ 1346 cells. In Section \ref{solidgasformation} we choose a planet with 0.1$\mathrm{M}_{\mathrm{earth}}$. The planet, fixed at 6\,au, accretes solids until it reaches its critical mass, when gas accretion begins. When the planet reaches a mass of about 30$\mathrm{M}_{\mathrm{earth}}$ there are 20 radial cells inside the planet's Hill region. Therefore, considering that the planet grows to 1$\mathrm{M}_{\mathrm{jup}}$ and comparing with the test performed above, the chosen resolution is adequate for our analyses.

In Section \ref{mig} the growing protoplanet is allowed to migrate, so the situation is more complicated. This is because the planet travels through regions of the gas disc with different densities and because the size of the Hill region changes with the planetary mass and position. So, to test if the chosen resolution is adequate in that case, we start with a low resolution and increase it until the results obtained for the final mass and for final position of the migrating planet differ in less than 5\%.
Future work will deal with a moving mesh that fits the region near the planet with a fixed number of cells, so it will be possible to analyze with more detail how the resolution influences the gas accretion model during planetary migration.


\bsp	
\label{lastpage}
\end{document}